\documentclass[aps,preprint,floats,tightenlines,nofootinbib,superscriptaddress,showpacs]{revtex4}

\usepackage{latexsym}
\usepackage{amsmath2000}
\usepackage{amssymb}
\usepackage{amsfonts}
\usepackage{amscd}

\numberwithin{equation}{section}

\def\A{{\ensuremath{\cal A}}}
\def\Ab{\ensuremath{\bar{\cal A}}}
\def\Ab{\ensuremath{\bar{\cal A}}}

\def\HA{\ensuremath{\cal{HA}}}
\def\H{{\ensuremath{\cal H}}}
\def\HS{\ensuremath{{\cal H}_{\mathrm{Sch}}}}
\def\HP{\ensuremath{{\cal H}_{\mathrm{Poly}}}}

\def\HPt{\ensuremath{{\cal H}}_{\mathrm{Poly}}}
\def\S{{\ensuremath{\cal S}}}

\def\cyl{{\mathrm{Cyl}}}
\def\cylstar{{\mathrm{Cyl}}^\star}
\def\cylt{{\mathrm{Cyl}}}
\def\cylstart{{\mathrm{Cyl}}^\star}
\def\SU(2){\mathrm{{SU(2)}}}
\def\U(1){\mathrm{{U(1)}}}
\def\a{\bar{A}}

\def\j{\ensuremath{{\vec j}}}
\def\n{\ensuremath{{\vec n}}}
\def\g{\ensuremath{\gamma}}
\def\gt{\ensuremath{\gamma}}
\def\ab{\ensuremath{\bar{A}}}

\def\u{\underline}
\def\be{\begin{equation}}
\def\ee{\end{equation}}
\def\ba{\begin{eqnarray}}
\def\ea{\end{eqnarray}}

\newcommand{\field}[1]{\ensuremath{\mathbb{#1}}}
\newcommand{\C}{\field{C}}
\newcommand{\R}{\field{R}}
\newcommand{\Z}{\field{Z}}
\newcommand{\Rbar}{\R}
\newcommand{\Cbar}{\C}

\newcommand{\rexpect}[3]{\ensuremath{(#1\,|\,#2\,|\,#3\rangle}}

\newcommand{\norm}[1]{\ensuremath{\|#1\|}}
\newcommand{\Hilb}{\ensuremath{\mathcal{H}}}
\newcommand{\bra}[1]{\ensuremath{\langle#1|}}
\newcommand{\ket}[1]{\ensuremath{|#1\rangle}}
\newcommand{\brar}[1]{\ensuremath{(#1|}}  
\newcommand{\ip}[2]{\ensuremath{\langle#1\,|\,#2\rangle}}

\newcommand{\evalue}[1]{\ensuremath{\langle#1\rangle}}

\newcommand{\Khl}{\ensuremath{\widehat{K^2_{\ell}}}}
\newcommand{\Khmu}{\ensuremath{\widehat{K^2_{\mu_o}}}}

\begin{document}

\title{Quantum gravity, shadow states, and quantum mechanics}

\author{Abhay Ashtekar}
\email{ashtekar@gravity.psu.edu}
\affiliation{Center for Gravitational Physics and Geometry\\
             Department of Physics, The Pennsylvania State
             University\\
             University Park, PA 16802-6300, USA \vspace{0.35em}}
\affiliation{Erwin Schr\"odinger Institute \\
             9 Boltzmanngasse \\
             1090 Vienna, Austria \vspace{0.35em}}

\author{Stephen Fairhurst}
\email{sfairhur@phys.ualberta.ca}
\affiliation{Center for Gravitational Physics and Geometry\\
             Department of Physics, The Pennsylvania State
             University\\
             University Park, PA 16802-6300, USA \vspace{0.35em}}
\affiliation{Erwin Schr\"odinger Institute \\
             9 Boltzmanngasse \\
             1090 Vienna, Austria \vspace{0.35em}}
\affiliation{Theoretical Physics Institute\\
             University of Alberta\\
             Edmonton, Alberta T6G 2J1, Canada}

\author{Joshua L. Willis}
\email{jwillis@gravity.psu.edu}
\affiliation{Center for Gravitational Physics and Geometry\\
             Department of Physics, The Pennsylvania State
             University\\
             University Park, PA 16802-6300, USA \vspace{0.35em}}

\begin{abstract}

A program was recently initiated to bridge the gap between the
Planck scale physics described by loop quantum gravity and the
familiar low energy world. We illustrate the conceptual problems
and their solutions through a toy model: quantum mechanics of a
point particle. The model can also serve as a simple introduction
to many of the ideas and constructions underlying quantum
geometry. Maxwell fields will be discussed in the second paper of
this series which further develops the program.

\end{abstract}

\pacs{04.60Pp, 04.60.Ds, 04.60.Nc, 03.65.Sq}
\maketitle

\section{Introduction}
\label{s1}

Perhaps the central conceptual lesson of general relativity is
\textit{gravity is geometry}. There is no longer a background
metric, no inert stage on which dynamics unfolds; like all
physical fields, geometry is dynamical. Therefore, one expects
that a fully satisfactory quantum gravity theory would also be
free of a background space-time geometry. However, of necessity, a
background independent description must use physical concepts and
mathematical tools that are quite different from those normally
used in low energy quantum physics. A major challenge, then, is to
show that this low energy description does arise from the
pristine, Planckian world in an appropriate sense. This challenge
is now being met step by step in the context of loop quantum
gravity. Some of the key ideas were summarized in \cite{al7},
which in turn was motivated by \cite{mv}. They will now be
discussed in detail and significantly extended in a series of
papers, of which this is the first (see also \cite{toh,st1}). Our
goal here is to illustrate, through a simple example, both the
tension between the two frameworks and the new physical notions
and mathematical techniques that are being used to resolve it.

Let us begin by listing some of the main issues and questions.

Loop quantum gravity is based on \textit{quantum geometry}, the
essential discreteness of which permeates all constructions and
results. The fundamental excitations are 1-dimensional and
polymer-like. A convenient basis of states is provided by spin
networks. Low energy physics, on the other hand, is based on
quantum field theories which are rooted in a flat space continuum.
The fundamental excitations of these fields are 3-dimensional,
typically representing wavy undulations on the background
Minkowskian geometry. The convenient Fock-basis is given by
specifying the occupation number in one particle states labelled
by momenta and helicities. At first sight, the two frameworks seem
disparate. Nonetheless, one would hope that the polymer
description admits semi-classical states which approximate
classical space-times as well as fluctuations on them represented
by gravitons and other fields. On the other hand, since this
perturbative description breaks down rather quickly because of
closed graviton loops, the low energy results are not likely to
emerge as first terms in a systematic expansion of a finite, full
theory. \textit{What then is the precise sense in which
semi-classical states are to arise from the full theory?}

{}From a mathematical physics perspective, the basic variables of
quantum geometry are holonomies (or Wilson loops) of the
gravitational connection $A$ along 1-dimensional curves and fluxes
of the conjugate momenta (the triads) $E$ across 2-surfaces. In
the final quantum theory, the connection $A$ fails to be a
well-defined operator(-valued distribution); only the holonomies
are well-defined. In perturbative quantum field theories, by
contrast, the vector potential operators are distributions,
whence, a priori, their holonomies fail to be well-defined
operators. Similarly, fluxes of electric field operators across
2-surfaces fail to be well-defined even on the Fock space of
photons. Heuristically, then, it would appear that, even at a
kinematic level, loop quantum gravity describes a `phase' of gauge
theories which is distinct from the one used in electrodynamics.
Since it is generally believed that distinct phases carry distinct
physics, it is natural to ask: \textit{Is the well-tested,
macroscopic `Coulomb phase' of low energy gravity compatible at
all with the Planck scale discreteness of quantum geometry?} If
so, in what sense? How does it emerge from loop quantum gravity?
Given the apparent deep differences, the procedure of extracting
the `Coulomb phase' from the fundamental Planckian description
should be rather subtle.

Finally, a further technical but important complication arises
from the detailed treatment of dynamics. Solutions to the quantum
Einstein equations (i.e. quantum constraints) do not belong to the
so-called kinematical Hilbert space $\HP$. This is not surprising:
a similar situation occurs already for simple, quantum mechanical
constrained systems. The kinematical Hilbert space provides the
mathematical framework to construct well-defined operators which
can be regarded as the quantum analogs of the classical constraint
functions. If zero lies in the continuous part of the spectrum of
these operators, none of the solutions to the quantum constraints
are normalizable with respect to the kinematic inner product.
(This is the case even for the simple constraint $p_x = 0$ in
$\Rbar^3$, and for the constraint $g^{ab} p_a p_b - \mu^2 =0$
satisfied by a free particle in Minkowski space-time.) The
solutions are distributional; they belong to the \textit{dual} of
a sub-space of `nice' quantum states (e.g. the Schwartz space).
The situation is completely analogous in quantum gravity. The
`nice' quantum states are typically taken to be finite linear
combinations of spin network states and their space is denoted by
$\cyl$ (the space of `\emph{cylindrical}' functions of
connections). Solutions to the quantum Einstein equations belong
to its dual, $\cylstar$. There is an inclusion relation (providing
a `Gel'fand-type' triplet) $\cyl \subset \HP \subset \cylstar$.
While the kinematical spin network states belong to $\cyl$, the
physical states belong to $\cylstar$. Therefore, semi-classical
states, capturing the low energy physics, should also be in
$\cylstar$. The problem is that, as of now, $\cylstar$ does not
have a physically justified inner product; a definite Hilbert
space structure is not yet available. Can one nonetheless hope to
extract low energy physics already at this stage? In particular,
\textit{can one test a candidate state in $\cylstar$ for
semi-classicality without access to expectation values?}

The primary purpose of this paper is to analyze these and related
issues using the simple example of a non-relativistic particle. We
will find that the issues raised above arise also in this example
and can be resolved satisfactorily. (For an analysis with similar
motivation, but which emphasizes the role of constraints and
discrete time evolutions, see \cite{bgp}.)

For readers who are not familiar with quantum geometry, this
example can also serve as an introduction to the mathematical
techniques used in that framework. However, as is typically the
case with toy models, one has to exercise some caution. First,
motivations behind various construction often become obscure from
the restrictive perspective of the toy model, whence the framework
can seem cumbersome if one's only goal is to describe a
non-relativistic particle. Secondly, even within mathematical
constructions, occasionally external elements have to be brought
in to mimic the situation in quantum geometry. Finally, because
the toy model fails to capture several essential features of
general relativity, there are some key differences between the
treatment of the Hamiltonian and other constraints in the full
theory and that of the Hamiltonian operator in the toy model. With
these caveats in mind, the toy model can be useful in
understanding the essential differences between our background
independent approach and the Fock-space approach used in
Minkowskian, perturbative quantum field theory.

We will begin with the usual Weyl algebra generated by the
exponentiated position and momentum operators. The standard
Schr\"{o}dinger representation of this algebra will play the role
of the Fock representation of low energy quantum field theories
and we will construct a new, unitarily inequivalent
representation---called the \textit{polymer particle
representation}---in which states are mathematically analogous to
the polymer-like excitations of quantum geometry. The mathematical
structure of this representation mimics various features of
quantum geometry quite well; in particular there are clear analogs
of holonomies of connections and fluxes of electric fields,
non-existence of connection operators, fundamental discreteness,
spin networks, and the spaces $\cyl$ and $\cylstar$%
\footnote{Of course, since this is only a simple, `toy example',
it does not capture all the subtleties. In particular, we will see
that a number of distinct notions in quantum geometry often
coalesce to a single notion in the example.}.
At the basic mathematical level, the two descriptions are quite
distinct and, indeed, appear to be disparate. Yet, we will show
that states in the standard Schr\"{o}dinger Hilbert space define
elements of the analog of $\cylstar$. As in quantum
geometry, the polymer particle $\cylstart$ does \textit{not} admit
a natural inner product. Nonetheless, as indicated in \cite{al7},
we \textit{can} extract the relevant physics from elements of
$\cylstart$ by examining their \textit{shadows}, which belong to
the  polymer particle Hilbert space $\HP$. This physics is
indistinguishable from that contained in Schr\"odinger quantum
mechanics in its domain of applicability.

These results will show that, in principle, one could adopt the
viewpoint that the polymer particle representation is the
`fundamental one'---it incorporates the underlying discreteness
of spatial geometry---and the standard Schr\"{o}dinger
representation corresponds only to the `coarse-grained' sector of
the fundamental theory in the continuum approximation. Indeed,
this viewpoint is viable from a purely mathematical physics
perspective, i.e., if the \textit{only} limitation of
Schr\"{o}dinger quantum mechanics were its failure to take into
account the discrete nature of the Riemannian geometry. In the
real world, however, the corrections to non-relativistic quantum
mechanics due to special relativity and quantum field theoretic
effects largely overwhelm the quantum geometry effects, whence the
above viewpoint is not physically tenable. Nonetheless, the
results for this toy model illustrate why an analogous viewpoint
can be viable in the full theory: Although the standard, low
energy quantum field theory seems disparate from quantum geometry,
it can arise, in a systematic way, as a suitable semi-classical
sector of loop quantum gravity.

The paper is organized as follows. Section \ref{s2} recalls a few
essential notions from quantum geometry which motivate our
construction of the polymer particle representation. This
representation is constructed in detail in section \ref{s3}. In
section \ref{s4} we show that the standard coherent states of the
Schr\"{o}dinger theory can be regarded as elements of $\cylstar$,
introduce the notion of `shadow states' and use them to show that
the elements of $\cylstar$ defined by the coherent states are, in
a precise sense, semi-classical from the perspective of the
`fundamental' polymer particle representation. In section \ref{s5}
we discuss dynamics in the polymer particle representation. To
define the kinetic energy term in the Hamiltonian, on can mimic
the procedure used to define the Hamiltonian constraint operator
in quantum general relativity. However, in the toy model, this
requires the introduction of a new structure by hand, namely a
fundamental length scale, which can be regarded as descending from
an underlying quantum geometry. The resulting dynamics is
indistinguishable from the standard Schr\"odinger mechanics in the
domain of applicability of the non-relativistic approximation.
Deviations arise only at energies which are sufficiently high to
probe the quantum geometry scale. In particular, shadows of the
Schr\"odinger energy eigenstates are excellent approximations  to
the `more fundamental' polymer eigenstates.

In the second paper in this series we will show that we can extend
these ideas to quantum field theory, where the familiar low energy
physics can be extracted from a more fundamental theory based on
quantum geometry. The two appendices of the present paper contain
some technical material which will be important to that analysis.

\section{Quantum Geometry}
\label{s2}

This summary of quantum geometry will enable the reader to see the
parallels between quantum geometry and the polymer particle
representation constructed in section \ref{s3}. It will be used
primarily to motivate our constructions in subsequent sections.
Our discussion will be rather brief and, in particular, we will
omit all proofs. (These can be found, e.g., in %
[6--17].) 
A detailed understanding of this material is not essential to the rest
of the paper. 

In diffeomorphism invariant theories of connections, the phase
space consists of pairs of fields $(A, E)$ on a 3-manifold
$\Sigma$, where $A_a^i$ are connection 1-forms which take values
in the Lie-algebra of the structure group $G$, and  $E^a_i$ are
`electric fields' which are vector densities with values in the
dual of the Lie algebra. For the purpose of this paper, it
suffices to restrict ourselves to two special cases: i) $G=
\SU(2)$, used in quantum geometry, and, ii) $G = \U(1)$ used in
quantum Maxwell theory. In either case, the `elementary' classical
observables are taken to be holonomies $A_e$ along paths $e$
defined by $A$ and fluxes $E_S$ of electric fields across
2-surfaces $S$. {}From the perspective of the standard Hamiltonian
formulation of field theories, these functions are `singular':
Since they are supported on 1-dimensional curves and 2-dimensional
surfaces, respectively, we are in effect using distributional
smearing functions. Nonetheless, the symplectic structure on the
classical phase space endows them with a natural Lie bracket and
the resulting Lie-algebra is taken as the point of departure in
quantum theory.

The Hilbert space of states can be constructed in two ways. In the
first, one uses the fact that, as usual, the configuration
variables $A_e$ give rise to an Abelian $C^\star$ algebra $\HA$,
called the holonomy algebra. One then introduces a natural
(diffeomorphism invariant) positive linear functional on it and
uses the Gel'fand-Naimark-Segal (GNS) construction to obtain a
Hilbert space $\HP$ of states and a representation of $\HA$ on it.
Finally, self-adjoint electric flux operators are introduced on
$\HP$ using the heuristic idea that $E$ should be represented by
$-i\hbar\,\delta/\delta A$.%
\footnote{{}From the viewpoint of the algebraic approach, which
has been so successful in quantum field theory in curved
space-times, working with a specific Hilbert space representation
may seem restrictive. However, the algebraic approach is not so
well-suited for systems, like general relativity, with non-trivial
constraints. More importantly, there is no loss of generality in
working with the above representation because it is singled out
essentially by the requirement of diffeomorphism covariance
\cite{st2}.}
The second approach is more explicit. One begins by specifying the
space $\cyl$ of `nice' functions of connections. Fix a graph $\g$
on the 3-manifold $\Sigma$ with $N$ edges. A connection $A$
associates to each edge $e$ a holonomy $A_e \in G$. The space of
$N$-tuples $(A_1, \ldots, A_N)$ defines a configuration of the
gauge theory restricted to the graph $\g$ and will be denoted by
$\A_\g$. Clearly, $\A_\g$ is isomorphic with $G^N$. Now, given a
smooth, complex-valued function $\psi$ on $G^N$, we can define a
function $\Psi$ of connections in an obvious fashion:
$$ \Psi(A) = \psi (A_{1}, \ldots, A_{N}). $$
The space of these functions is denoted $\cyl_\g$. Elements of
$\cyl_\g$ have knowledge only of the connection \textit{restricted
to} $\g$. The space $\cyl$ of \textit{all} cylindrical functions
is obtained by simply considering all possible graphs $\g$:
$$  \cyl = \bigcup_\g \cyl_\g .   $$
Thus, each element of $\cyl$ depends only on holonomies of the
connection along edges of \textit{some} finite graph $\g$ but the
graph can vary from one function to another. Had we restricted
ourselves to a fixed graph $\g$, the theory would have been
equivalent to a lattice gauge theory on a (generically irregular)
`lattice' $\g$. However, since we allow \textit{all} possible
graphs, we are dealing with a \textit{field} theory, with an
infinite number of degrees of freedom, of all connections on
$\Sigma$.

The next step is to introduce an inner product on $\cyl$. For
this, we simply use the induced Haar measure $\mu_H^{(N)}$ on
$\A_\g \approx G^N$: Given any two functions $\Psi_1$ and $\Psi_2$
on $\cyl_\g$, we set
\be \label{ip1} (\Psi_1, \Psi_2) = \int_{\A_\g} \, \bar{\psi_1}\,
\psi_2 \,\, d\mu_H^{(N)}  .\ee
Using properties of the Haar measure one can verify that this
definition is unambiguous, i.e., if $\Psi_1$ and $\Psi_2$ are
cylindrical with respect to another graph $\g'$, the right side of
(\ref{ip1}) is unchanged if we replace $\g$ with $\g^\prime$. This
prescription provides us with an Hermitian inner product on all of
$\cyl$ because, given any $\Psi_1, \Psi_2 \, \in \cyl$, there
exists a (sufficiently large) graph $\g$ such that $\Psi_1, \Psi_2
\in \cyl_\g$. The Cauchy completion of $\cyl$ with respect to this
inner product provides the required Hilbert space $\HP$ of all
quantum states, obtained in the first method via the GNS
construction.

Because we consider \textit{all} possible graphs on $\Sigma$ in
its construction, $\HP$ is very large. However, it can be
decomposed into convenient \textit{finite} dimensional sub-spaces.
Each of these subspaces is associated with a labelling of edges of
a graph $\g$ by non-trivial irreducible representations of $G$.
Thus, in the case when $G= \SU(2)$, let us label each edge $e$ of
$\g$ with a non-zero half-integer (i.e., spin) $j_e$. Then, there
is a finite dimensional sub-space $\H_{\g, \j}$ such that
\be \label{sndecomp} \HP = \bigoplus_{\g, \j} \, \, \H_{\g, \j}.
\ee
This is called the \textit{spin network decomposition} of $\HP$.
Although $\HP$ is very large, practical calculations are feasible
because each of the sub-spaces $\H_{\g, \j}$ can be identified
with the Hilbert space of a spin-system which is extremely well
understood. In the case when $G= \U(1)$, we label each edge $e$
with a non-zero integer $n_e$. The Hilbert space $\H_{\g, \n}$ is
now \textit{1-dimensional}, spanned by the function
$$ \Psi(A) = e^{in_1\theta_1}\cdots e^{in_N\theta_N} $$
where $e^{i\theta_m}$ is the holonomy of the connection $A$ along
the edge $e_m$. These functions are called \textit{flux network
states} and by replacing ${\vec j}$ by ${\vec n}$ in
(\ref{sndecomp}) one now obtains a decomposition of $\HP$ in terms
of 1-dimensional orthonormal subspaces.

As in any Schr\"odinger description, quantum states in $\HP$ can
be regarded as square integrable function on the quantum
configuration space. In systems with finite number of degrees of
freedom, the quantum configuration space is normally the same as
the classical one. However, for systems with an infinite number of
degrees of freedom, there is typically a significant enlargement:
while classical configurations are represented by smooth fields,
quantum configurations are \textit{distributional}. This occurs
also in our case: $\HP = L^2(\Ab, d\mu_o)$, where $\Ab$ is a
suitable completion of the space $\A$ of smooth connections and
$\mu_o$, a regular measure on it. An element $\ab$ of $\Ab$ is
called a \textit{generalized connection}. It associates with every
oriented path $e$ in $\Sigma$ an element $\ab(e)$ of $G$, the
holonomy along $e$ subject only to two conditions: i)
$\ab(e_1\circ e_2) = \ab(e_1)\, \ab(e_2)$; and, ii) $\ab(e^{-1}) =
[\ab(e)]^{-1}$. Note that the assignment $e \longrightarrow \a(e)$
can be \textit{arbitrarily discontinuous}, whence the quantum
configuration space $\Ab$ is a genuine extension of the classical
configuration space $\A$. Nonetheless, in a natural topology, $\A$
is dense in $\Ab$, whence $\Ab$ can be regarded as a suitable
completion of $\A$. However, as is typically the case in field
theories, the measure $\mu_o$ is concentrated on genuinely
generalized connections; all the smooth configurations in $\A$ are
contained in a set of zero measure.

The measure $\mu_o$ is completely defined by the family of
measures $\mu_H^{(N)}$ on $\A_\g \approx G^N$: because
$\mu_H^{(N)}$ are mutually consistent in a precise sense, they can
be `glued together' to obtain $\mu_o$. Indeed, \textit{every}
measure on $\Ab$ arises as a \textit{consistent} family of
measures on $\A_\g$. More generally, structures in the full
quantum theory are constructed as \textit{consistent} families of
structures on $\A_\g$ or $\cyl_\g$. In particular, many of the
physically interesting operators on $\HP$---such as the holonomies
$\hat{A}_e$, the fluxes $\hat{E}_S$ of $\hat{E}$ across $S$, area
operators $\hat{A}_S$ associated with 2-surfaces $S$, and volume
operators $\hat{V}_R$ associated with spatial regions $R$---arise as
consistent families of operators on $\cyl_\g$.
Therefore, their properties can be explored in terms of their
actions on finite dimensional spaces $\H_{\g, \j}$ (or $\H_{\g,
\n}$).

While the above structures suffice to discuss quantum kinematics,
as pointed out in the Introduction, an additional notion is needed
in the discussion of quantum dynamics: solutions to the quantum
Einstein's equations do not belong $\H$ because they fail to be
normalizable. Their natural home is $\cylstar$, the algebraic dual
of $\cyl$. We have a natural inclusion:
 $$ \cyl \subset \HP \subset \cylstar . $$
To discuss physical states and explore the physically relevant
semi-classical sector, then, we are led to focus on $\cylstar$.

We will see in section \ref{s3} that the essential features of
these constructions and results are mirrored in a transparent way
in the `polymer particle representation' of a non-relativistic
point particle.

\section{Schr\"odinger and Polymer Particle Frameworks}
\label{s3}

The physical system we wish to consider is a particle moving on
the real line $\Rbar$. (It is straightforward to extend our
discussion to $\Rbar^n$.) A natural point of departure for quantum
theory is provided by the Weyl-Heisenberg algebra. To each complex
number $\zeta$ associate an operator $W(\zeta)$ (which will turn
out to be a product of the exponentiated position and momentum
operators) and consider the free vector space $\mathbf{W}$
generated by them. Introduce a product on $\mathbf{W}$ via:
\be \label{wp1}
 W(\zeta_1) W(\zeta_2) = e^{\frac{i}{2}\,
 {\mathrm{Im}}\zeta_1 \bar{\zeta_2}}\,\,
 W(\zeta_1 + \zeta_2),
\ee
and an involution $\star$ via
\be [W(\zeta)]^\star = W(-{\zeta})\, . \ee
This is the Weyl-Heisenberg $\star$-algebra of non-relativistic
quantum mechanics. Here, as in the mathematical literature, we
have chosen $\zeta$ to be dimensionless.

In the physics literature, one implicitly introduces a length
scale $d$ and `splits' operators $W(\zeta)$ by setting
$$ W(\zeta) = e^{{\frac{i}{2} \lambda \mu}}\, U(\lambda)\,
V(\mu) $$
where $\zeta = \lambda d+ i(\mu/d)$. Thus, $U(\lambda) = W(\lambda d)$
and $V(\mu) = W(i\mu/d)$. The operators $U(\lambda)$ and $V(\mu)$ satisfy
$[U(\lambda)]^\star = U(-\lambda),\,\, [V(\mu)]^\star = V(-\mu)$ and
the product rule:
\begin{gather}
 \label{wp2}
 U(\lambda_1)U(\lambda_2) =
 U(\lambda_1+\lambda_2),\quad
 V(\mu_1) V(\mu_2) = V(\mu_1 +\mu_2), \notag\\
  U(\lambda)V(\mu) = e^{-i \lambda \mu}\, V(\mu)\, U(\lambda) \, .
\end{gather}
Therefore, in any representation of $\mathbf{W}$, the Hilbert space
carries  1-parameter unitary groups $U(\lambda),\, V(\mu)$. To fix
notation and make a detailed comparison, we will first recall the
standard Schr\"odinger representation of this algebra and then
introduce the polymer particle representation in some detail.

\subsection{The Schr\"odinger representation}
\label{s3.1}

The celebrated Stone-von Neumann  theorem ensures us that every
irreducible representation of $\mathbf{W}$ \textit{which is weakly
continuous in the parameter} $\zeta$ is unitarily equivalent to
the standard Schr\"odinger representation, where the Hilbert space
is the space $L^2(\Rbar, d\u{x})$ of square integrable functions
on $\Rbar$ (where $\u{x}$ is dimensionless). $W(\zeta)$ are
represented via:
\be
 \hat{W}(\zeta) \psi(\u{x}) = e^{\frac{i}{2} \alpha\beta}\,
 e^{i\alpha \u{x}} \,\, \psi(\u{x}+\beta),
\ee
where $\zeta = \alpha + i\beta$. This is an irreducible
representation of $\mathbf{W}$. Furthermore, the $\hat{W}(\zeta)$
are all unitary (i.e., satisfy $[\hat{W}(\zeta)]^\dagger =
[\hat{W}(\zeta)]^{-1}$) and weakly continuous in $\zeta$ (i.e.,
all matrix elements of $\hat{W}(\zeta)$ are continuous in
$\zeta$).

In physics terms, the Hilbert space $\HS$ is the space of square
integrable functions of $x = \u{x}d$ and the action of these
operators is given by
\be \hat{U}(\lambda)\, \psi(x) = e^{i\lambda x}\, \psi(x) \quad
\text{and} \quad \hat{V}(\mu)\, \psi(x) = \psi(x+ {\mu}) \ee
for all $\psi \in \HS$. Now, the 1-parameter unitary groups
$\hat{U}(\lambda)$ and $\hat{V}(\mu)$ are weakly continuous in the
parameters $\lambda, \mu$. This ensures that there exist
self-adjoint operators $\hat{x}$ and $\hat{p}$ on $\HS$ such that
\be\label{sch1}
    \hat{U}(\lambda) : = e^{i \lambda \hat{x}} \qquad \mbox{and}
    \qquad \hat{V}(\mu) = e^{i\frac{\mu}{\hbar}\, \hat{p}} \, .
\ee

We conclude with two remarks:

\noindent i) The Schr\"{o}dinger representation can be obtained
using the Gel'fand-Naimark-Segal (GNS) construction with the
positive linear (or, `expectation-value') functional $F_{\mathrm{Sch}}$
on $\mathbf{W}$:
\be F_{\mathrm{Sch}} (W(\zeta)) = e^{-\frac{1}{2}|\zeta|^2}. \ee
The expectation values of $\hat{U}$ and $\hat{V}$ are given by:
\be \label{sch2} F_{\mathrm{Sch}} (U(\lambda)) = e^{-\frac{1}{2}
\lambda^2 d^2} \qquad \text{and} \qquad F_{\mathrm{Sch}} (V(\mu)) =
e^{-\frac{1}{2} \frac{\mu^2}{d^2}} .\ee
The corresponding GNS `vacuum' (i.e., cyclic) state
$\psi_{\mathrm{Sch}}$ is simply
$$\psi_{\mathrm{Sch}} (x) = (\pi d^2)^{-\textstyle{\frac{1}{4}}}\,\,
e^{-\textstyle{\frac{x^2}{2d^2}}}\, .$$

\noindent ii) For definiteness, we have presented the
Schr\"{o}dinger representation using position wave
functions $\psi(x)$. In terms of momentum wave functions $\psi(k)$,
which will be more useful in the next subsection, we have:
\begin{equation}
\label{eq:schruv}
\hat{U}(\lambda) \psi (k) = \psi (k-\lambda), \quad \text{and}
\quad \hat{V}(\mu) \psi (k) = e^{i\mu k}\, \psi(k)
\end{equation}
and the GNS cyclic state is given by:
$$ \psi_{\mathrm{Sch}} (k) = \left(\frac{\pi}{ d^2}
\right)^{-\textstyle{\frac{1}{4}}}\,\,
e^{-\textstyle{\frac{k^2d^2}{2}}}.$$

\subsection{The polymer particle representation}
\label{s3.2}

We are now ready to introduce the polymer particle representation
of the Weyl-Heisenberg algebra which is unitarily
\textit{inequivalent} to the Schr\"odinger. This construction
must, of course, violate one or more assumptions of the Stone-von
Neumann uniqueness theorem. It turns out that only one assumption
is violated: in the new representation, the operator $V(\mu)$ will
not be weakly continuous in $\mu$, whence there will be no
self-adjoint operator $\hat{p}$ such that $V(\mu) = \exp{(i\mu
\hat{p})}$. While the unavailability of the standard momentum
operator seems alarming at first, this is just what one would
expect physically in the absence of a spatial continuum. More
precisely, if the spatial Riemannian geometry is to be discrete
(as, for example, in loop quantum gravity), one would \textit{not}
expect the operator $p = -i\hbar\,d/dx$ to exist at a fundamental
level. The key question is whether one can nonetheless do quantum
mechanics and reproduce the well-tested results. This is a
difficult question with many technical subtleties. But, as we will
see in sections \ref{s4} and \ref{s5}, the answer is in the
affirmative: by adopting the viewpoint that the natural arena for
quantum theory is the analog of $\cylstar$, one can recover
results of Schr\"odinger quantum mechanics in the domain of its
validity.

To bring out the similarity and differences with quantum geometry,
we will construct the Hilbert space of states, $\HPt$, in steps,
using the same terminology. A \emph{graph} $\g$ will consist of a
countable set $\{ x_i \}$ of points on the real line $\Rbar$ with
the following two properties: i) The $x_i$ do not contain
sequences with accumulation points in $\Rbar$; and, ii) there
exist constants $\ell_\g$ and $\rho_\g$ such that the number
$n(I)$ of points in any interval $I$ of length $\ell(I) \ge
\ell_\g$ is bounded by $n(I) \le \rho_\g \ell(I)$. The two
technical conditions will ensure convergence of certain series;
see section \ref{s4}.%
\footnote{In the earlier version of this paper, we only had
condition i). We thank Jacob Yngvasson for pointing out that it
does not suffice and Chris Fewster and Jerzy Lewandowski for the
precise formulation of ii).}

Denote by $\cylt_{\gt}$ the space of complex valued functions
$f(k)$ of the type:
\be \label{cylt} f(k) = \sum_j f_j \, e^{-ix_j k}  \ee
on $\Rbar$, where $x_j$ are real and $f_j$ are complex numbers
with a suitable fall-off. To simplify the later specification of
domains of operators, we will choose the fall-off to be such that
$\sum_j |x_j|^{2n}|f_j|^2 < \infty$ for all $n$. $\cylt{}_{\gt}$
is a vector space (which is infinite dimensional if the number of
points in $\gt$ is infinite). We will say that functions $f(k)$ in
$\cylt{}_{\gt}$ are \textit{cylindrical} with respect to $\gt$.
Thus, each cylindrical state is a \textit{discrete} sum of plane
waves; it fails to belong to the Schr\"odinger Hilbert space. The
real number $k$ is the analog of connections in quantum geometry
and the plane wave $\exp{(-ikx_j)}$ can be thought of as the
`holonomy of the connection $k$ along the edge $x_j$'.

Next, let us consider all possible graphs, where the points (and
even their number) can vary from one graph to another, and denote
by $\cylt$ the \textit{infinite} dimensional vector space of
functions on $\Rbar$ which are cylindrical with respect to
\textit{some} graph. Thus, any element $f(k)$ of $\cylt$ can be
expanded as in (\ref{cylt}), where the uncountable basis $\exp{(-ix_j
  k)}$ is now labeled by \textit{arbitrary} real numbers $x_j$.
Let us introduce a natural, Hermitian inner product on $\cylt$ by
demanding that $\exp{(-ix_j k)}$ are orthonormal:
\be \label{innerprod} <e^{-ix_i k} | e^{-ix_jk} > \, = \,
\delta_{x_i, x_j}. \ee
(Note that the right side is the Kronecker $\delta$ and not the
Dirac distribution.) Denote by $\HPt$ the Cauchy completion of
$\cylt$. This is the Hilbert space underlying our representation.

To summarize, $\HPt$ is the Hilbert space spanned by countable
linear combinations $\sum_1^\infty f_{j} \exp{(-ix_j k)}$ of plane
waves in the momentum space, subject to the condition
$\sum_1^\infty |f_j|^2 < \infty$, where $\{ x_j\}$ is an arbitrary
countable set of real numbers, which can vary from one state to
another. Even more succinctly, $\HPt = L^2(\Rbar_{\mathrm{d}},
d\mu_{\mathrm{d}})$, where $\Rbar_{\mathrm{d}}$ is the real line equipped
with discrete topology and $\mu_{\mathrm{d}}$ is the natural discrete
measure on it.

The Weyl-Heisenberg algebra $\mathbf{W}$ is represented on $\HPt$ in
the same manner as in the Schr\"odinger representation:
\be \hat{W}(\zeta) f(k) = [e^{\frac{i}{2} {\lambda\mu}}\,\,
U(\lambda)\, V(\mu)]\, f(k) \ee
where, as before, $\zeta= \lambda d + i (\mu/d)$ and the action of
$\hat{U}$ and $\hat{V}$ is given by (see~(\ref{eq:schruv}))
\be \hat{U}(\lambda) f(k) = f(k-\lambda) \quad \text{and} \quad \hat{V}(\mu)
f(k) = e^{{i\mu k}} \,\, f(k). \ee
It is straightforward to check that these operators provide a
faithful, irreducible representation of $\mathbf{W}$ on $\HPt$. Each
$\hat{U}(\lambda)$ and $\hat{V}(\mu)$ is unitary.

The structure of this representation becomes more transparent in
terms of eigenkets of $\hat{U}(\lambda)$. Let us associate with the
basis elements $\exp{(-ix_j k)}$ a ket $\ket{x_j}$ and, using the
textbook heuristic notation, express $\exp{(-ix_j k)}$ as a
generalized scalar product:
$$ (k, {x_j}) = e^{-ix_j k}  $$
Then, $\{ \, \ket{x_j}\,  \}$ is an orthonormal basis and the
action of the basic operators $\hat{U}$ and $\hat{V}$ is given by:
\be \label{uvops} \hat{U}(\lambda) \ket{x_j} \, =\, e^{i\lambda
x_j} \ket{x_j} \quad \text{and} \quad \hat{V}(\mu) \ket{x_j} \, =
\, \ket{x_j - \mu}. \ee

It is straightforward to verify that $\hat{U}(\lambda)$ is weakly
continuous in $\lambda$ whence there exists a self-adjoint
operator $\hat{x}$ on $\HPt$ with $\hat{U}(\lambda) = \exp
(i\lambda \hat{x})$. Its action can now be expressed as:
\be \label{xop} \hat{x}\ket{x_j} \, = \, x_j \ket{x_j}  \ee
just as one would expect. However, there is an important
difference from the Schr\"odinger representation: The eigenkets of
$\hat{x}$ are now \textit{normalizable}, and hence elements of the
Hilbert space itself. In this sense the eigenvalues are
`discrete'.

By contrast, although the family $\hat{V}(\mu)$ provides a
1-parameter unitary group on $\HPt$, it \textit{fails to be weakly
continuous} in the parameter $\mu$. This follows from the fact
that, no matter how small $\mu$ is, $\ket{x_j}$ and $\hat{V}(\mu)
\ket{x_j}$ are orthogonal to one another, whence
$$\lim_{{\mu}\mapsto 0} \bra{x_j} \hat{V}(\mu) \ket{x_j}
\, = 0 \, ,$$
while $\hat{V}(\mu=0) = 1$ and $\ip{x_j}{x_j}\, = \, 1$. Thus,
there is no self-adjoint operator $\hat{p}$ on $\HPt$ satisfying
the second of eqs.~(\ref{sch1}).

Finally, this representation can be obtained via
Gel'fand-Naimark-Segal construction, using the following positive
linear (or expectation value) functional on the Weyl-Heisenberg
algebra $\mathbf{W}$:
\be F_{\mathrm{Poly}}(W(\zeta)) =\begin{cases} 1 &
\text{if $\mathrm{Im}\,\zeta = 0$,} \\
0 & \text{otherwise.} \end{cases} \ee
In terms of $U(\lambda)$ and $V(\mu)$, we have:
\begin{align}
F_{\mathrm{Poly}}(U(\lambda)) &=  1 \qquad \forall\,\lambda, \notag \\
F_{\mathrm{Poly}}(V(\mu)) &=
      \begin{cases}
        1 & \text{if $\mu = 0$,} \\
        0 & \text{otherwise.}
      \end{cases}
\end{align}
The corresponding cyclic state is simply $\ket{\psi_o}\, = \,
\ket{x_{o}\!=\!0}$. Note that, in contrast to the Schr\"odinger
positive linear functional $F_{\mathrm{Sch}}$, no scale had to be
introduced in the definition of $F_{\mathrm{Poly}}$. This is the analog
of the fact that the corresponding positive linear functional in
quantum geometry is diffeomorphism invariant.

We will conclude with a few remarks.

\noindent i) The step by step procedure used above brings out the
fact that the polymer particle description captures many of the
mathematical features of quantum geometry, but now for a very
simple physical system. Our notation is geared to reflect the
analogies. Thus, sets $\gt= \{ x_k \}$ are the analogs of graphs
of quantum geometry; individual points $x_j$, the analogs of
edges; the continuous, momentum variable $k$, the analog of
connections; $\exp{(-ix_j k)}$ the analog of the holonomy along an
edge; $\cylt_{\gt}$ the analog of the space of cylindrical
functions associated with a graph and $\cyl$ the space of all
cylindrical functions of quantum geometry; and the $\ket{x_j}$ the
analogs of spin network states. Indeed, we again have the Hilbert
space decomposition analogous to (\ref{sndecomp}):
$$ \HPt = \bigoplus_{x} \H_x $$
where $\H_x$ are the 1-dimensional subspaces spanned by our basis
vectors $\ket{x}$. (The decomposition is thus analogous to that in
the $\U(1)$ case).

\noindent ii) What is the situation with operators? The basic
operators of quantum geometry---holonomies and fluxes of the
electric field---are respectively analogous to the operators
$\hat{V}(\mu)$ and $\hat{x}$ on $\HPt$. The commutator between
$\hat{x}$ and $\hat{V}(\mu)$,
\begin{equation}\label{commute}
   [\hat{x}, \hat{V}(\mu)] = - \mu \hat{V}(\mu),
\end{equation}
is completely parallel to the commutator between electric fields
and holonomies in quantum geometry. Just as $\hat{V}(\mu)$ are unitary
but $\hat{p}$ does not exist, holonomies are unitarily represented
but the connection operator does not exist. Like $\hat{x}$, the
electric flux operators are unbounded self-adjoint operators with
discrete eigenvalues. (However, in the case of electric fluxes,
the set of eigenvalues is a discrete subset of the real line,
equipped with its standard topology.) It is this discreteness that
leads to the loss of continuum in the quantum Riemannian geometry
which in turn `justifies' the absence of the standard momentum
operator $-i\hbar\,d/dx$ in the polymer particle example.

\noindent iii) Recall that in quantum geometry, elements of $\HP$
can be represented as \textit{functions} on a compact space $\Ab$,
the quantum configuration space obtained by a suitable completion
of the classical configuration space $\A$. What is the situation
with respect to $\HPt$? Now, the classical configuration space is
just the real line $\Rbar$ (of momenta $k$). The quantum
configuration space turns out to be the Bohr compactification
$\bar{\Rbar}_{\mathrm{Bohr}}$ of $\Rbar$ (discovered and analyzed
by the mathematician Harald Bohr, Niels' brother). All quantum
states are represented by square integrable functions on
$\bar{\Rbar}_{\mathrm{Bohr}}$ with respect to a natural measure
$\mu_o$; $\HPt = L^2( \bar{\Rbar}_{\mathrm{Bohr}}, d\mu_o)$.
Finally, as in quantum geometry, $\bar{\Rbar}_{\mathrm{Bohr}}$ is
also the Gel'fand spectrum of the \textit{Abelian}
$C^\star$-algebra of `holonomy' operators $V(\mu)$. (For details
on the Bohr compactification, see \cite{bohr}.)

\section{Relation between Schr\"{o}dinger and Polymer Descriptions:
Kinematics} \label{s4}

Elements of the polymer Hilbert space $\HPt$ consist of
\textit{discrete} sums $f(k) = \sum_j f_j\,\exp{(-ix_jk)} $ of plane
waves. Therefore, it follows that the intersection of $\HPt$ with
$\HS$ consists just of the zero element. While each provides an
irreducible, unitary representation of the Weyl-Heisenberg
algebra, the two Hilbert spaces are `orthogonal'. Therefore, one
might first think that the standard physics contained in the
Schr\"odinger representation cannot be recovered from the polymer
framework. We will now show that this is \textit{not} the case.

As explained in the introduction, the key idea is to focus on
$\cylstart$, the algebraic dual%
\footnote{As in quantum geometry, we are taking the algebraic dual
just for simplicity. When the framework is further developed, one
would introduce an appropriate topology on $\cyl$ (which is finer
than that of $\HPt$) and define $\cylstar$ as the space of linear
functions on $\cyl$ which are continuous in this topology. The
algebraic dual is `too large' but this fact is not relevant here:
since our main goal is to represent all semi-classical
Schr\"odinger states by elements of $\cylstart$ we can just ignore
the fact that the algebraic dual also contains other `unwanted'
states.}
of $\cylt$. Since $\cylt \subset \HPt$, it follows that we have:
$$ \cylt \subset \HPt \subset \cylstart \, . $$
We will denote the elements of $\cylstart$ by upper case letters,
e.g.,  $(\Psi|$, and their action on elements $\ket{f}$ of $\cylt$
simply with a juxtaposition, e.g. $(\Psi|$ maps $\ket{f}$ to the
complex number $(\Psi \ket{f}$.

The Weyl-Heisenberg algebra has a well-defined action on $\cyl$,
and hence by duality, on $\cyl^\star$:
\be \left[(\Psi| \hat{W}(\zeta)\right]\,\ket{f} = (\Psi|\,
\left[(\hat{W}(\zeta))^\dagger \ket{f}\right] \ee
However, this representation is far from being irreducible. In
particular, $\HPt$ is contained in $\cylstart$ and provides us
with an irreducible representation. More importantly for what
follows, the Schwartz space $\S$, a dense subspace of $\HS$
consisting of smooth functions on $\Rbar$ which, together with all
their derivatives fall off faster than any inverse polynomial in
$x$, is also embedded in $\cylstart$. (This follows from the two
technical conditions in the definition of a graph and, of course,
the definition of $\cyl$.) Since all coherent states belong to
$\S$ and they form an over-complete basis in $\HS$, Schr\"odinger
quantum mechanics is somehow encoded in $\cylstart$. Our task is
to analyze this encoding.

We will often use the fact that $\S$ is stable under Fourier
transform; i.e., $\psi(x) \in \S$ if and only if its Fourier
transform $\tilde{\psi}(k)\in \S$. The embedding of $\S$ in
$\cylstart$ is given just by the Schr\"odinger scalar product:
each element $\psi \in \S$ defines an element $(\Psi|$ of
$\cylstart$ via
\be \label{image} (\Psi|\,\left[\sum_j \, f_j\,\ket{e^{-ix_j k}}\right] =
\frac{1}{\sqrt{2\pi}} \, \sum_j \, f_j \int dk
\,\tilde{\bar{\psi}}(k) e^{-ix_jk} = \sum_j f_j\, \bar{\psi}(x_j)
\ee
where $\tilde{\psi}(k)$ is the Fourier transform of $\psi(x)$.
Thus, although elements of $\cylt$ fail to be normalizable in the
Schr\"odinger Hilbert space, their Schr\"odinger inner product
with elements of $\S$ is well-defined and naturally leads to a
linear map from $\cylt$ to $\Cbar$.

Can we exploit the fact that $\S$ is embedded in $\cylstart$ to
extract the physics of Schr\"odinger quantum mechanics from
$\cylstart$? At first sight, there appears to be a key problem:
$\cylstart$ is not equipped with a scalar product. We could
restrict ourselves just to $\S \subset \cylstart$ and introduce on
it the Schr\"odinger scalar product by hand. But this would just
be an unnecessarily complicated way of arriving at the
Schr\"odinger representation. More importantly, in
non-perturbative quantum gravity, we do not have the analog of the
Schr\"odinger Hilbert space and, furthermore, indications are that
its perturbative substitute, the graviton Fock space, is `too
small'. Therefore, for our polymer particle toy model to be an
effective tool, we should not restrict ourselves to a `small'
subspace of it such as $\S$. Rather, we should work with the full
$\cylstart$ and use only that structure which is naturally
available on it. Thus, our challenge is to show that standard
quantum physics can be extracted from $\cylstart$ directly,
without making an appeal to the Schr\"odinger Hilbert space. Known
facts about the Schr\"odinger representation can be used only to
motivate various constructions, but not in the constructions
themselves.

In quantum gravity, a principal open problem is that of existence
of semi-classical states. Therefore, in the rest of this section
we will focus on the problem of isolating elements of $\cylstart$
which correspond to the standard coherent states of Schr\"odinger
quantum mechanics and extracting their physics using only those
structures which are naturally available in the polymer framework.
Hamiltonians and their various properties will be discussed in the
next section.

\subsection{Isolating semi-classical states}
\label{s4.1}

Fix a classical state, i.e., a point $(x_o, p_o)$ in the classical
phase space. In Schr\"odinger quantum mechanics, the corresponding
semi-classical states are generally represented by coherent states
peaked at this point. In these states, the product of
uncertainties in the basic observables $\hat{x}$ and $\hat{p}$ is
minimized, $(\Delta\, \hat{x})\, (\Delta\, \hat{p}) = \hbar/2$,
and furthermore, in suitable units, these uncertainties are
distributed `equally' among the two observables. To obtain a
specific coherent state, one has to specify these units, or, in
physical terms, `tolerance' ---the uncertainties in $x$ and $p$ we
can tolerate. Let us therefore introduce a length scale $d$ and
ask that the uncertainty $\Delta\, x$ in $\hat{x}$ be $d/\sqrt{2}$
and that in $\hat{p}$ be $\hbar/(\sqrt{2}\,d )$. (In the case of
an harmonic oscillator, $d$ is generally taken to be
$\sqrt{\hbar/m\omega}$. However, in this section on kinematics, it
is not necessary to restrict ourselves to a specific system.) Set
$$\zeta_o = \frac{1}{\sqrt{2}d} \left(x_o + i \frac{d^2}{\hbar}p_o\right)
 = \frac{1}{\sqrt{2}d} \left(x_o + i {k _o}d^2\right)$$
where, from now on, we will use $k_o := p_o/\hbar$. Then, the
standard coherent state $\psi_{\zeta_o}$ is generally obtained by
solving the eigenvalue equation
\be \label{coh1} \hat{a}\, \psi_{\zeta_o} (x) \equiv
\frac{1}{\sqrt{2}\, d} \left(\hat{x} + i \frac{d^2}{\hbar}\hat{p}\right)\,
\psi_{\zeta_o} (x) = {\zeta_o}\, \psi_o (x), \ee
whose solution is
\be \label{cohstate1} \psi_{\zeta_o}(x) = c\,
e^{-\frac{(x-x_o)^2}{2 d^2}}\,\, e^{ik_o(x-x_o)},\ee
where $\hat{a}$ is the annihilation operator and $c$ is a
normalization constant. Since $\psi_{\zeta_o} \in \S$, it
canonically defines an element $\Psi_{\zeta_o}$ of $\cylstart$.
Our first task is to isolate this $\Psi_{\zeta_o}$ using just the
polymer framework. The second task, that of analyzing its
properties and specifying the sense in which it is a
semi-classical state also from the polymer perspective, will be
taken up in the next subsection.

Now, in the polymer framework, the operator $\hat{p}$ fails to be
well-defined. Therefore, we can not introduce the creation and
annihilation operators used in the above construction. However,
recall that the operators $\hat{U}(\lambda)$, $\hat{V}(\mu)$ and
$\hat{x}$ \textit{are} well-defined on $\cylt$ and hence also on
$\cylstart$. We can therefore reformulate (\ref{coh1}) by an
equivalent eigenvalue equation in terms of these operators. Since
the equation is now to be imposed on $\cylstart$, we have to
replace the annihilation operator $\hat{a}$ by its adjoint,
$\hat{a}^\dagger$, the creation operator. Now, using the
Baker-Hausdorff-Campbell identity in $\HS$, we have:
$$ e^{\sqrt{2}\alpha \hat{a}^\dagger} = e^{\frac{\alpha}{d}\hat{x}}
 \, V(-\alpha d) \, e^{-\frac{\alpha^2}{2}}.
$$
where the factor of $\sqrt{2}$ is introduced just for technical
simplification and $\alpha$ is an arbitrary real number. Note that
the operators on the right side are all well-defined on
$\cylstart$.

Collecting these ideas motivated by results in the Schr\"odinger
representation, we are now led to seek the analog of coherent
states in $\cylstart$ by solving the eigenvalue equation:
\be \label{coh2} (\Psi_{\zeta_o}|\,\left[e^{\frac{\alpha}{d}\hat{x}} \,
V(-\alpha d) \, e^{-\frac{\alpha^2}{2}}\right]\,  = \,
e^{\sqrt{2}\alpha\,\bar{\zeta_o}}\, (\Psi_{\zeta_o}|. \ee
for all real numbers $\alpha$. Note that, to capture the full
content of the original eigenvalue equation (\ref{coh1}), it is
essential to allow arbitrary $\alpha$ in the exponentiated version
(\ref{coh2}).

To obtain the solution, it is convenient to use a basis in
$\cylstart$. Recall first that any element $f$ of $\cylt$ can be
expanded out as a discrete sum, $f = \sum_j f_j\,\ket{x_j}$, where
the $f_j$ are complex coefficients and the $x_j$ real numbers. Therefore,
the action of any element $(\Psi|$ of $\cylstart$ is completely
determined by the action $(\Psi\ket{x} = \overline\Psi(x)$ of $(\Psi|$
on all basis vectors $\ket{x}$. That is, $(\Psi|$ can be expanded
as a continuous sum
\be \label{Psi} (\Psi| = \sum_x \, \overline\Psi(x) (x|  \ee
where the dual basis $(x|$ in $\cylstart$, labeled by real numbers
$x$, is defined in an obvious fashion:
$$ (x\ket{x_j}\, = \, \delta_{x, x_j}.  $$
Note that, although there is a continuous sum in (\ref{Psi}),
when operating on any element of $\cyl$ only a countable number
of terms are non-zero.

Using (\ref{Psi}) in (\ref{coh2}), it is straightforward to show
that the coefficients $\Psi_{\zeta_o}(x)$ must satisfy:
\be \label{coh3} \overline\Psi_{\zeta_o}(x+ \alpha d)\, = \,\exp{
\left[\sqrt{2}\alpha \bar{\zeta}_o - \frac{\alpha x}{d} +
\frac{\alpha^2}{2}\right]} \, \overline\Psi(x) \ee
for all real numbers $\alpha$. It is easy to verify that this
equation admits a solution which is unique up to a normalization
factor. The general solution is given by:
\be \label{cohstate2} \brar{\Psi_{\zeta_{o}}} = \bar{c}\,\,
\sum_x\, \left[e^{- \frac{(x-x_o)^2}{2d^2}}\,\,
e^{-ik_o(x-x_o)}\right] \, (x|\, .\ee
As one might have hoped, the coefficients in this expansion are
the same as the expression (\ref{cohstate1}) of the coherent state
wave function in the Schr\"odinger representation. Note that, to
obtain a unique solution (up to a multiplicative constant), it is
essential to impose (\ref{coh3}) for \textit{all} real numbers
$\alpha$.

To summarize, by using the standard procedure in the Schr\"odinger
representation as motivation, we wrote down an eigenvalue equation
directly in $\cylstart$ to single out a candidate semi-classical
state $(\Psi_{\zeta_o}|$ peaked at a generic point $(x_o, p_o)$ of
the classical phase space. Since this is a linear equation, one
cannot hope to restrict the overall normalization of the solution.
Up to this trivial ambiguity, however, the solution is unique. We
will refer to it as a \textit{polymer coherent state}. As one
might have hoped, this polymer coherent state is just the element
$(\Psi_{\zeta_o}|$ of $\cylstart$ defined by the standard coherent
state $\psi_{\zeta_o} \in \S$ in $\HS$. Note that this is not an
assumption but the result of a self-contained calculation that was
carried out entirely in $\cylstart$. However, at this stage, it is
not a priori obvious that $(\Psi_{\zeta_o}|$ is a semi-classical
state \textit{from the polymer perspective}, especially because we no
longer have access to the Schr\"odinger scalar product. This issue
will be discussed in the next subsection.

\subsection{Shadow States}
\label{s4.2}

For simplicity, in this subsection we will restrict ourselves to
the candidate semi-classical state $(\Psi_o|$ corresponding to
$\zeta =0$. (The general case is completely analogous and
discussed in subsection \ref{s4.3}.) Our task is to show that this
state is sharply peaked at $x$=0 and $p$=0 using only the polymer
framework. However, right at the outset we encounter two
difficulties. Firstly, the operator $\hat{p}$ is not defined in
the polymer framework. We will therefore have to define a
`fundamental operator' on $\HP$ which is approximated by $\hat{p}$
of the Schr\"odinger representation. The second difficulty is
that, since there is no inner product on $\cylstart$, the required
expectation values cannot be defined on it. To overcome this
obstacle, we will use graphs as `probes' to extract physical
information from elements $(\Psi|$ of $\cylstart$. More precisely,
we will `project' each $(\Psi|$ to an element
$\ket{\Psi_{\g}^{\mathrm{shad}}}$ in $\cylt_\gamma$ and analyze
properties of $(\Psi| $ in terms of its shadows
$\ket{\Psi_{\g}^{\mathrm{shad}}}$. Each shadow captures only a
part of the information contained in our state, but the collection
of shadows can be used to determine the properties of the full
state in $\cylstart$.

Let us begin by defining the required projection $\hat{P}_{\gt}$
from $\cylstart$ to $\cylt_{\gt}$:
\be (\Psi|\,\hat{P}_{\gt} := \sum_{x_j\in\gt}\Psi(x_j) \,\ket{x_j}
\, \equiv \ket{\Psi^{\mathrm{shad}}_\gamma}\, . \ee
The ket $\ket{\Psi^{\mathrm{shad}}_\gamma}$ defines the shadow
cast by the element $\brar{\Psi}$ of $\cylstart$ on the graph
$\gamma$ in the sense that
$$ (\Psi\ket{f_\g} = \ip{\Psi^{\mathrm{shad}}_\gamma}{f_\g} $$
where the left side is the result of the action of an element of
$\cylstart$ on an arbitrary element $f_\g$ of $\cylt_\g$ and the
right side is the scalar product on $\cylt_\g$. Our task is to
analyze properties of the shadows
$$\ket{\Psi_{o,\gamma}^{\mathrm{shad}}} := (\Psi_{o}|\,\hat{P}_{\gt}
 \, . $$
of our candidate semi-classical state. The essential idea is to
say that $(\Psi_o|$ is semi-classical if physical observables of
interest have expected mean values with small uncertainties in its
shadows \ket{\Psi_{o,\gamma}^{\mathrm{shad}}} on sufficiently
refined graphs $\gamma$.

To make this notion precise, we need to select: i) A suitable
family of graphs; ii) a class of observables of interest; and,
iii) acceptable `tolerances' for mean-values and uncertainties of
these observables. We will restrict ourselves to shadows on
\emph{regular} lattices
\footnote{Quantum geometry considerations imply that, to probe
semi-classicality, we should only use those graphs in which the
number of points in \emph{any} macroscopic interval is
proportional to the length of the interval. Regular lattices offer
the simplest way to achieve this. A priori one may be concerned
that this is `too small a class'. But the results of this section
show that it suffices.}
with sufficiently small lattice spacing (as discussed below). For
definiteness, as in Schr\"odinger quantum mechanics, the class
${\cal C}$ of observables of interest will consist just of
position and momentum operators. Tolerances $\tau$ will be
determined by the physical parameters of the system under
consideration (i.e., the length scale $d$ of subsection
\ref{s4.1}).

We will say that a state $(\Psi| \in \cylstart$ is semi-classical
with respect to ${\cal C}$ and peaked at a point $(x,p)$ of the
classical phase space, if within specified tolerances $\tau$, the
`expectation values' of any operator $\hat{A} \in {\cal C}$ equals
the classical value $A(x,p)$ and the fluctuations are small; i.e.,
if
\be
\label{semiclass}
\frac{(\Psi|\hat{A}\ket{\Psi_\g^{\mathrm{shad}}}}
{\norm{{\Psi_\g^{\mathrm{shad}}}}^2} \, = A(x,p)(1 + \tau^{(1)}_A)
\,\,\, {\rm and}\,\,\,
\frac{(\Psi|\hat{A^2}\ket{\Psi_\g^{\mathrm{shad}}}}
{\norm{{\Psi_\g^{\mathrm{shad}}}}^2} \,\, - \,\,
\left(\frac{(\Psi|\hat{A}\ket{\Psi_\g^{\mathrm{shad}}}}
{\norm{{\Psi_\g^{\mathrm{shad}}}}^2}\right)^2  \le \tau^{(2)}_A \ee
for all sufficiently refined graphs $\g$. Here $\norm{f}$ is the
norm of the state $\ket{f}$ in $\HPt$, and $\tau_A^{(i)}$ are the
tolerances assigned to the observable $A$. The meaning of the
equation is clearer if the operators are thought as acting on the
candidate semi-classical state $(\Psi|$ in $\cylstar$ by duality.
Thus, in the first equality, the `expectation value' of $\hat{A}$
in the candidate semi-classical state $(\psi|$ is evaluated by the
action of $(\Psi|\hat{A}$ ($\in \cylstar$) on the shadow
$\ket{\Psi_\g^{\mathrm shad}}$ of $(\Psi|$ on the graph $\gamma$.
If the action of the operator $\hat{A}$ leaves $\cyl_\g$
invariant, as one might hope, this `expectation value' reduces to
the more familiar expression $\bra{\Psi_\g^{\mathrm
shad}}\hat{A}\ket{\Psi_\g^{\mathrm shad}}$. However, for more
general operators, the two expressions do not agree and
$(\Psi|\hat{A}\ket{\Psi_\g^{\mathrm{shad}}}$ turns out to be the
better measure of the expectation value.

Let us then work with infinite regular lattices with spacing
$\ell$, where $\ell$ is chosen to be sufficiently small (see
below). The shadow of our candidate semi-classical state
$(\Psi_o|$ on the regular graph is given by:
\begin{equation}\label{psil}
    \ket{\Psi_{o,\ell}^{\mathrm{shad}}} = c\,\,
     \sum_{n\in \Z} e^{-\textstyle{\frac{n^{2}\ell^{2}}{2d^{2}}}}\,
     \ket{n\ell} \, ,
\end{equation}
where $c$ is an arbitrary constant. We can now compute the
expectation values and fluctuations of various operators in detail
and examine if the state can be regarded as semi-classical. On
general grounds, one would hope to obtain good agreement with the
standard coherent state of Schr\"odinger quantum mechanics
provided the lattice spacing $\ell$ is much smaller than the
length scale $d$ that defines our tolerance. We will show that,
although there are subtleties, this expectation is borne out.
However, let us first pause to examine whether this requirement is
physically reasonable.  As an example, consider the vibrational
oscillations of a carbon monoxide molecule. These are well
described by a harmonic oscillator with parameters
$$ m \approx 10^{-26}\,\mathrm{kg} \quad \text{and} \quad \omega
\approx   10^{15}\,\mathrm{Hz} $$
The textbook treatment of the harmonic oscillator implies that we
cannot require the tolerance $d$ for $\hat{x}$ to be smaller than
 $$ d_{\mathrm{min}} = \sqrt{\frac{\hbar}{m\omega}}
 \approx 10^{-12}\, \mathrm{m}\, ;$$
if we did, the resulting state would spread out quickly under
quantum evolution. On the other hand, since no evidence of spatial
discreteness has been observed at particle accelerators, the
quantum geometry viewpoint requires us to choose  $\ell <
10^{-19}\mathrm{m}$, and we may even wish to move $\ell$ all the
way down to the Planck scale ($\ell_{p} = 1.6\times 10^{-35} \,
\mathrm{m}$). Thus, our assumption that $\ell \ll d$ is well
justified. Working in this regime, we will now show that the
quantities computed using (\ref{psil}) agree to leading order with
the standard Schr\"odinger coherent state. The corrections are of
order $\ell^2/d^2 < 10^{-14}$ and, furthermore, appear in the
regime in which Schr\"odinger quantum mechanics is inapplicable
due to, e.g.,  relativistic effects.

Let us begin with the norm of the shadow of the polymer coherent
state:
\begin{equation}\label{norm0}
\ip{\Psi_{o,\ell}^{\mathrm{shad}}}{\Psi_{o,\ell}^{\mathrm{shad}}}
\,= \, |c|^2\,\, \sum_{n = -\infty}^{\infty}
e^{-\textstyle{\frac{n^2\ell^2}{d^2}}} \, .
\end{equation}
Here, we have used the fact that $\ip{x_i}{x_j} =
\delta_{x_{i},x_{j}}$ to simplify the double sum to a single one.
Now, since $\ell \ll d$, the exponential on the right hand side of
(\ref{norm0}) decays very slowly, whence we can not estimate the
norm by keeping just a few terms in the sum. Fortunately, however,
we can use the Poisson re-summation formula:
\begin{equation}\label{poisson}
   \sum_{n} g(x+n) = \sum_{n = -\infty}^{\infty}  e^{2\pi i\, x \,n}
   \int_{-\infty}^{\infty} g(y) e^{-2\pi i \, y\, n} dy \, ,
\end{equation}
for all functions $g(y)$ which are suitably well behaved for the
sums to converge. We will take
$$ g(y) = e^{-\textstyle{\frac{y^2 \ell^2}{d^2}}} \quad \mbox{and} \quad
x = 0 \, . $$
Then it is straightforward to calculate
\begin{equation}\label{norm1}
    \ip{\Psi_{o,\ell}^{\mathrm{shad}}}{\Psi_{o,\ell}^{\mathrm{shad}}}\,
    =\, |c|^2\, \frac{\sqrt{\pi}
    d}{\ell}\,\, \sum_{n = -\infty}^{\infty}
    e^{-\textstyle{\frac{\pi^{2}n^{2}d^2}{\ell^{2}}}}
    \approx \, |c|^2\, \frac{\sqrt{\pi} d}{\ell}
    \left(1 + 2 e^{-\textstyle{\frac{\pi^{2}d^2}{\ell^{2}}}}
    \right)\, ,
\end{equation}
where we have used  $(d/\ell)\gg 1$ to truncate the series after
the second term.

Next we turn to the expectation value of and fluctuations in
$\hat{x}$. For semi-classicality, the expectation value should be
close to zero and the fluctuations of the order $d/\sqrt{2}$. For
expectation values, we obtain
\begin{equation}\label{x}
   \brar{\Psi_{o}}\hat{x} |\Psi_{o,\ell}^{\mathrm{shad}}\rangle\,
   =\,|c|^2\, \sum_{n} (n\ell) \,
   e^{-\textstyle{\frac{n^{2}\ell^2}{d^2}}} = 0 \, ,
\end{equation}
due to antisymmetry in $n$. This result agrees exactly with that
obtained from the Schr\"odinger coherent state. Let us turn to the
calculation of fluctuations. We have
\begin{eqnarray}\label{xsq}
   \brar{\Psi_{o}}\hat{x}^{2} |\Psi_{o,\ell}^{\mathrm{shad}}\rangle
   &=& |c|^2\, \sum_{n} (n\ell)^2 \,
   e^{-\textstyle{\frac{n^{2}\ell^2}{d^2}}} \nonumber\\
   &=& |c|^2\, \frac{\sqrt{\pi}d^3}{2\ell} \sum_{n}
   e^{-\textstyle{\frac{\pi^2 n^2 d^2}{\ell^2}}} \left(1 -
   \frac{2\pi^{2}n^2 d^2}{\ell^{2}} \right)\, ,
\end{eqnarray}
where we have once again made use of the Poisson re-summation
formula.  By combining the results of (\ref{xsq}) and
(\ref{norm1}), we can obtain the fluctuations in $\hat{x}$,
\begin{equation}\label{deltax}
    (\Delta x)^{2} := \frac{\brar{\Psi_{o}} \hat{x}^{2}
    |\Psi_{o,\ell}^{\mathrm{shad}}\rangle}
    {\norm{\Psi_{o,\ell}^{\mathrm{shad}}}^2} -
    \left(\frac{\brar{\Psi_{o}}\hat{x}
    |\Psi_{o,\ell}^{\mathrm{shad}}\rangle}
    {\norm{\Psi_{o,\ell}^{\mathrm{shad}}}^2}\right)^2
    \approx \frac{d^2}{2} \left(1 - \frac{4\pi^{2}d^2}{\ell^{2}} \,
    e^{-\textstyle{\frac{\pi^{2}d^2}{\ell^{2}}}} \right) \, ,
\end{equation}
where we have made use of the fact that the expectation value of
$\hat{x}$ is zero.  Hence, we see that the fluctuations in
$\hat{x}$ satisfy our `tolerance' requirement. Indeed, to leading
order, they agree with the those in the standard coherent states
of the Schr\"odinger framework and the sub-leading terms are
extremely small, going to zero as $\ell/d$ tends to zero.
Interestingly, these corrections actually \textit{decrease} the
uncertainty in $x$ for the discrete case.

Thus, we see that our candidate semi-classical state $(\Psi_o|$ is
indeed sharply peaked at $x =0$. What about the momentum?  As
mentioned above, there is no natural analog of the Schr\"odinger
momentum operator $\hat{p}$ on $\HPt$. Thus, the viewpoint is that
the standard $\hat{p}$ operator is a `low energy' construct. There
are several operators in the `fundamental description' whose action
on `low lying states' is approximated by $\hat{p}$. Here, we will
choose one and test for semi-classicality of $(\Psi_o|$. As one
might hope, the difference between candidate choices is manifest
only at such high energies that the Schr\"odinger quantum
mechanics is inapplicable there.

To define an analog of the Schr\"odinger momentum operator, we
will use a standard strategy from lattice gauge theories. We first
note that, classically, if $k\mu$ is small then we can expand
$\exp(-ik\mu)$ as
\begin{equation}
   \exp(-ik\mu) = 1 - ik\mu - \frac{k^2 \mu^2}{2} + \cdots
\end{equation}
whence
\begin{equation}
   \frac{\exp(-ik\mu) - \exp(ik\mu)}{-2i\mu} = k + O(k^2\mu) \, .
\end{equation}
In quantum theory, then, it seems natural to define the analog of
the momentum operator in a similar way. Choose a sufficiently
small value $\mu_o$ of $\mu$ (with $\ell\le\mu_o \ll d)$ and
define the momentum operator on $\HPt$ as $\hat{p} =
\hbar\hat{K}_{\mu_o}$, with
\begin{equation}\label{K}
   \hat{K}_{\mu_o} := \frac{i}{2{\mu_o}} \left(\hat{V}(\mu_o) -
   \hat{V}(-\mu_o)\right) \, .
\end{equation}
(The simpler definition $\hat{K}_{\mu_o} = (i/2\mu_o)
(\hat{V}({\mu_o}) - 1)$ is not viable because this operator fails
to be self-adjoint.) With this definition in hand, let us examine
the expectation value and fluctuations in $\hat{K}_{\mu_o}$.
$(\Psi|$ will be semi-classical also for momentum if the
expectation value of $\hat{K}_{\mu_o}$ is close to zero and the
fluctuation is of the order $1/\sqrt{2} d$.

Now, a direct calculation in the polymer Hilbert space yields
\begin{equation}\label{vmu}
\langle \hat{V}(\mu) \rangle := \frac{\brar{\Psi_{o}} \hat{V}(\mu)
|\Psi_{o,\ell}^{\mathrm{shad}}\rangle}
{\norm{\Psi_{o,\ell}^{\mathrm{shad}}}^2}  \approx
e^{-\textstyle{\frac{\mu^2}{4d^2}}} \left( 1 + 2\,
e^{-\textstyle{\frac{\pi^{2}d^2}{\ell^{2}}}}
\left[\cos\left(\frac{\pi\mu}{\ell}\right) - 1 \right]\right) \, ,
\end{equation}
for any value of $\mu$. Using this result, it is
straightforward to show that
\begin{equation}
   \langle\hat{K}_{\mu_o} \rangle = 0
\end{equation}
because of the antisymmetry between $\hat{V}({\mu_o})$ and
$\hat{V}(-{\mu_o})$ in our definition (\ref{K}). Next, let us
analyze the fluctuations
\be  \langle \hat{K}_{\mu_o}^2\rangle = \frac{1}{4{\mu_o}^2}
\left(2 - \langle \hat{V}(2{\mu_o})\rangle -  \langle
\hat{V}(-2{\mu_o})\rangle \right)\, . \ee
Substituting $\mu = \pm 2\mu_{o}$ in the above expression
(\ref{vmu}), we obtain
\be \langle \hat{K}_{\mu_o}^2\rangle \, \approx \, \frac{1}{2{\mu_o}^2}
\left[1 - e^{-\textstyle{\frac{{\mu_o}^2}{d^2}}} \right] \,
   \approx\,\frac{1}{2d^2}\left[1 - \left(\frac{{\mu_o}^2}{2d^2}\right)
      \right]\, ,
\ee
where we have used the fact ${\mu_o} \ll d$ to expand in powers of
$({\mu_o}/d)$ in the last step.  Recalling that the expectation
value of $\hat{K}$ in the state $\ket{\Psi_{o}^{\ell}}$ is zero,
we obtain the fluctuations in $\hat{K}$ as
\begin{equation}\label{deltak}
   (\Delta K_{\mu_o})^2 \approx  \frac{1}{2d^2}\left[1 -
   \left(\frac{{\mu_o}^2}{2d^2}\right) \right] \, .
\end{equation}
Since the approximate momentum operator is given by
$\hbar\hat{K}_{\mu_o}$, we conclude that the state is sharply
peaked at $p= 0$ and the fluctuations are within the specified
tolerance.

Finally, collecting the results for $\Delta\, x$ and $\Delta\, k$,
we obtain the uncertainty relations for the shadow of the polymer
semi-classical state:
\begin{equation}\label{uncertxk}
   (\Delta x)^2 (\Delta K_{\mu_o})^2 = \frac{1}{4} \left[1 -
   \left(\frac{{\mu_o}^2}{2d^2}\right) +
   \mathcal{O}\!\left(\frac{{\mu_o}^4}{d^4}\right)  \right] \, .
\end{equation}
Note that the corrections to the standard uncertainty relations at
order $({\mu_o}/d)^2$ \textit{decrease} the uncertainty. This can
occur because the commutator between the position and the
approximate momentum operator is not simply a multiple of
identity. Such modifications of the uncertainty relations have
also been obtained in string theory. Our discussion shows that the
effect is rather generic.

To summarize, in subsection \ref{s4.1}, we found candidate
semi-classical states $(\Psi_{\zeta_o}|$ in $\cylstart$ working
entirely in the polymer particle framework. In this sub-section,
we showed that the polymer coherent state $(\Psi_o|$ is
semi-classical in the polymer sense: its shadows on sufficiently
refined lattices are sharply peaked at the point ($x$=0, $p$=0) of
the classical phase space. Furthermore, the fluctuations in $x$
and $p$ are essentially the same as those in the Schr\"odinger
coherent state $\psi_o$ of (\ref{cohstate1}). There \textit{are}
deviations, but in the regime of applicability of Schr\"odinger
quantum mechanics, they are too small to violate experimental
bounds.

\subsection{General coherent states}
\label{s4.3}

Let us now analyze the properties of general polymer coherent
states $(\Psi_{\zeta}|$ with $\zeta = \frac{1}{\sqrt{2}d}
\left(x+id^2 k \right)$. Calculations of expectation values and
fluctuations proceed in a very similar manner to those described
above for $(\Psi_o|$. (The only difference arises from the fact
that we may not have a point in our graph at the position $x$.
However, this affects the sub-leading terms.)  Therefore, we will
simply state the final results:

\begin{enumerate}

\item The norm of the state is given by
\begin{equation}
   \ip{\Psi_{\zeta,\ell}^{\mathrm{shad}}}
   {\Psi_{\zeta,\ell}^{\mathrm{shad}}} =
   |c|^2 \, \frac{\sqrt{\pi} d}{\ell}\,\, \left(
   1 + \mathcal{O}\!\left(e^{-\textstyle{\frac{\pi^2 d^2}
   {\ell^2}}}\right)\right) \, .
\end{equation}

\item The expectation value of the $\hat{x}$ operator is
\begin{equation}
   \langle\hat{x}\rangle\, := \,
   \frac{\rexpect{\Psi_{\zeta}}{\hat{x}}{\Psi_{\zeta,\ell}^{\mathrm{shad}}}}
   {\norm{\Psi_{\zeta,\ell}^{\mathrm{shad}}}^{2}} = x \, \left[1 +
   \mathcal{O}\!\left(e^{-\textstyle{\frac{\pi^2
   d^2}{\ell^2}}}\right)\right] \, .
\end{equation}
Thus, the expectation value of position is $x$ within the
tolerance $\tau^{(1)}_x = e^{-\textstyle{\frac{\pi^2
d^2}{\ell^2}}}$.

\item The fluctuation in $x$ is
\begin{equation}
   (\Delta x)^2 = \frac{d^2}{2}\, \left[1 +
   \mathcal{O}\!\left(e^{-\textstyle{\frac{\pi^2 d^2}{\ell^2}}}
   \right) \right] \, .
\end{equation}
So, the leading term, $d/\sqrt{2}$, in the fluctuation in $x$ is
the same as in the Schr\"odinger coherent states. Also, the
sub-leading terms are independent of $\zeta$, i.e., are the same
for all polymer coherent states.

\item One can evaluate the $\hat{K}_{\mu_o}$ operator on an arbitrary
coherent state. The result is,
\begin{equation}
   \evalue{\hat{K}_{\mu_o}} = k \, \left(1 + {\mathcal{O}}(k^2\mu_o^2) +
   \mathcal{O}\!\left(\frac{\ell^2}{d^2}\right) \right) .
\end{equation}
Thus, we now encounter a new situation. The tolerance
$\tau^{(1)}_{K_{\mu_o}}$ is acceptably small only if $k\mu_o \ll
1$. In this case, we obtain an uncertainty relation similar to the
one in (\ref{uncertxk}). However, for $k\mu_o \sim 1$ our states
do not satisfy the semi-classicality requirement. But note that
the non-relativistic approximation ---and hence the motivation for
including $\hat{K}_{\mu_o}$ in the list ${\cal C}$ of
observables--- breaks down long before one reaches such high
momenta. (In the case of the CO molecule, for example, this would
correspond to the energy level $n \ge 10^{14}$.)

\end{enumerate}

To summarize, we have introduced polymer coherent states
$(\Psi_\zeta|$ and investigated their properties using their
shadows $\ket{\Psi_\zeta^\ell}$. Given a tolerance $d$ for
$\hat{x}$, an uniform graph can serve as a suitable `probe'
provided the lattice spacing $\ell$ is chosen so that $\ell/d \ll
1$. As far as semi-classical states are concerned, systems which
can be treated adequately within non-relativistic quantum
mechanics can also be well-described by the polymer particle
framework, without any reference to the Schr\"odinger Hilbert
space.

\emph{Remark}: Recall that the normalized Schr\"odinger coherent
states $\ket{\psi_\zeta}$ form an overcomplete basis in $\HS$
providing a convenient resolution of the identity:
\begin{equation} \label{oc1}
   \int dk \int dx\,\ket{\psi_{\zeta}}
   \bra{\psi_{\zeta}}\, = \, I\, .
\end{equation}
Does a similar result hold for the shadow coherent states
$\ket{\Psi_{\zeta,\ell}^{\mathrm{shad}}}$ in the Hilbert space
$\HPt^\ell$ restricted to the graph? A priori, it would appear
that there is a potential problem. Since
$$\ket{\Psi_{\zeta,\ell}^{\mathrm{shad}}} = c \sum_n
\left(e^{(n\ell -x)^2}\, e^{ik(n\ell -x)}\right)\,\, \ket{n\ell}$$
where $\zeta = \frac{1}{\sqrt{2}d}(x + id^2k)$, it follows that
the projection operators
$$P_{\zeta} := \frac{\ket{\Psi_{\zeta,\ell}^{\mathrm{shad}}}
   \bra{\Psi_{\zeta,\ell}^{\mathrm{shad}}}}
   {\norm{\Psi_{\zeta,\ell}^{\mathrm{shad}}}^2} $$
defined by the shadow coherent states are periodic: $P_{\zeta} =
P_{\zeta^\prime}$ where $k^\prime = k + (2\pi N)/\ell$. Thus,
while the label $k$ took values on the entire real line in
(\ref{oc1}), with shadow coherent states in $\HPt^\ell$, it can
only take values in $[-\pi/\ell, \pi/\ell]$. Therefore, one might
be concerned that, because of the `effective momentum cut-off' we
may not have `sufficient' coherent states for the standard
over-completeness to hold. However, it turns out that this concern
is misplaced. $\HPt^\ell$ is sufficiently small because of the
restriction to a fixed lattice for an \textit{exact}
over-completeness of the desired type to hold \cite{krp,gd} :
\begin{equation}
   \int_{-\pi/\ell}^{\pi/\ell} \frac{dk}{2\pi}
   \int_{-\infty}^{\infty}  dx \, P_{\zeta}
   =\, \sum_{n} \ket{n\ell}\,\bra{n\ell} \, =\, I_\ell\, ,
\end{equation}
where $I_\ell$ is the identity operator on $\HPt^{\ell}$.

\section{Relation between Schr\"{o}dinger and Polymer Descriptions:
Hamiltonians} \label{s5}

\subsection{The conceptual setting}
\label{s5.1}

Since a secondary goal of this paper is to illustrate strategies
used in loop quantum gravity, let us begin by recalling the
situation with the Hamiltonian constraint of quantum general
relativity \cite{ham}. The main term in the classical constraint
is of the form ${\rm Tr}\, E^a E^b F_{ab}$, where, as explained in
section \ref{s2}, the triad fields $E$ are the analogs of $x$ in
the polymer particle example and $F_{ab}$ is the curvature of the
gravitational connection $A$, the analog of $k$. While $E$'s and
holonomies of $A$ are well-defined operators on the quantum
geometry Hilbert space, connections are not. Therefore, $F_{ab}$
has to be expressed in terms of holonomies. Given a spin network
state, at each vertex, one introduces new edges, creating `small'
loops and expresses $F_{ab}$ in terms of holonomies along these
small loops (taking care of appropriate `area factors'). The
resulting operator initially depends on the choice these new
edges. However, while acting on \emph{diffeomorphism invariant
states (in $\cyl^\star$), the dependence on the details of these
edges drops out}. Thus, on states of physical interest, the final
Hamiltonian constraint does not make explicit reference to details
such as the lengths and `positions' of the new edges.

Let us now turn to the polymer particle. Now, the classical
Hamiltonian is of the form
\begin{equation}\label{ham}
   H = \frac{p^2}{2m} + V(x) \, ,
\end{equation}
where $V(x)$ is a potential. Since the operator $\hat{x}$ is
well-defined in the polymer framework, the main technical problem
is that of defining the operator analog of $p^2$. Thus the
situation is analogous to that with the Hamiltonian constraint,
described above. Again, we will need to introduce some extra
structure (which is invisible classically), this time to define
the analog of $p^2$ in terms of `holonomies' $V(\mu)$ of the
`connection' $k$ on the full Hilbert space $\HPt$. {}From a
mathematical viewpoint, the obvious choice is an `elementary
length' $\mu_o$. Physically, this is motivated by the expectation
that such a scale will be provided by a deeper theory (such as
quantum geometry) through an underlying discreteness. \emph{{}From
now on, we will adopt the viewpoint that this discreteness is
fundamental, whence {observationally} only those $V(\mu)$ are
relevant for which $\mu = N\mu_o$, for an integer $N$.}

Given $\mu_o$, we will set
\be \label{h}  \hat{H} = \frac{\hbar^2}{2m}\widehat{K_{\mu_o}^2} +
V(\hat{x}) \quad {\rm where} \quad \Khmu = \frac{1}{\mu_{o}^2}\, (2-
V(\mu_o) - V(-\mu_o))\ee
Unfortunately, in this example, we do not have a useful analog of
the diffeomorphism invariance of loop quantum gravity which can
help remove the dependence on this extra structure. Therefore, the
final Hamiltonian operator on $\HPt$ will continue to depended on
$\mu_o$; the reference to the additional structure does not go
away. This is simply a consequence of the fact that a toy model
can not mimic all aspects of the richer, more complicated theory,
whence, to carry out constructions which are analogous to those in
the full theory, certain structures have to be introduced
`externally'. However, we will see that, if one chooses the
discreteness scale $\mu_o < 10^{-19}m$ as in Section \ref{s4.2},
in the domain of validity of non-relativistic quantum mechanics,
predictions derived from (\ref{h}) are indistinguishable from
those of Schr\"odinger quantum mechanics and therefore in
agreement with experiments. \emph{In contrast to results of
section} \ref{s4}, \emph{this holds for fully quantum mechanical
results, not just the semi-classical ones.} At first sight, this
may seem obvious. However, the detailed analysis will reveal that
certain subtleties arise and have to be handled carefully. These
issues provide concrete hints for the precise procedure required
to compare the polymer and continuum theories in the much more
complicated context of quantum geometry. Thus, while the toy model
is constructed to mimic the situation in the full theory, its
concrete results, in turn, provide guidance for the full theory.

Since the key difficulties in the polymer description involve the
kinetic term, to illustrate the similarity and differences between
the polymer and Schr\"odinger dynamics it suffices to work with a
fixed potential. To facilitate the detailed comparison, in this
paper we will focus on the harmonic oscillator potential. (Results
on general potentials will be discussed elsewhere \cite{jw}.)

\emph{Remarks}: i) In the semi-classical considerations of the
last section, we had to find a `fundamental' operator on $\HPt$
which is the analog of the Schr\"odinger momentum operator.
Technically, the situation with the kinetic term in the
Hamiltonian, discussed above, is completely analogous. However,
there is a conceptual difference: whereas the operator
$\hat{K}_{\mu_o}$ was used only for semi-classical purposes,
$\hat{H}$ is to govern `fundamental dynamics' on $\HPt$.
Therefore, it has to be constructed and analyzed more carefully.
In particular, $\Khmu \not= \hat{K}_{\mu_o}^2$; we will see that
the latter choice gives an unwanted degeneracy in the eigenvalues
of $\hat{H}$.\\
ii) Since the \emph{final} Hamiltonian now depends on $\mu_o$, in
the polymer description, $\mu_o$ now has a fundamental
significance. This strengthens the viewpoint that the algebra of
physical observables is generated only by $\hat{V}(N\mu_o)$ and
$\hat{x}$.

\subsection{Eigenvalues and eigenstates of $\hat{H}$ in $\HPt$}
\label{s5.2}

Recall that a general element $\ket\Psi$ of $\HPt$ can be expanded
out as $\ket{\Psi} = \sum_x \Psi(x) \ket{x}$ (where $\Psi(x)$ is
non-zero only at a countable set of points). Therefore, the
eigenvalue equation $\hat{H} \ket{\Psi} =E \ket{\Psi}$ becomes a
difference equation on the coefficients $\Psi(x)$:
\be \label{ev1} \Psi(x+\mu_o) + \Psi(x- \mu_o) = \left[ 2-
\frac{2E}{\hbar\omega} \frac{\mu_o^2}{d^2} +
\frac{x^2}{d^2}\frac{\mu_o^2}{d^2} \right] \, \Psi(x)\, . \ee
The form of this equation implies that a basis of solutions is
given by states of the type
$$\ket{\Psi_{x_o}} = \sum_{m = -\infty}^{\infty}
\Psi_{x_o}^{(m)}\ket{x_o+ m\mu_o} \, \in \cyl_{\alpha^{x_o}} \,
,$$
where $\alpha^{x_o}$ is the regular lattice consisting of points
$x_o+m\mu_o$ with $x_o \in [0,\mu_o)$. For these states, the
difference equation reduces to a recursion relation
\be \label{ev2} \Psi_{x_o}^{(m+1)} + \Psi^{(m-1)}_{x_o} = \left[
2- \frac{2E}{\hbar\omega} \frac{\mu_o^2}{d^2}+\frac{(x_o +
m\mu_o)^2} {d^2}\frac{\mu_o^2}{d^2} \right] \, \Psi^{(m)}_{x_o}\,
. \ee
The full polymer Hilbert space $\HP$ can be decomposed as a direct
sum of separable Hilbert spaces $\HP^{x_o}$,
$$ \HP = \bigoplus_{x_o \in [0,\mu_o)}\, \HP^{x_o}\, ,$$
and the above energy eigenstates belong to the sub-space
$\HP^{x_o}$ of $\HP$. Note that since the observable algebra is
now generated by $\hat{x}$ and $\hat{V}(N\mu_o)$, observables can
not mix states belonging to distinct $\HP^{x_o}$; each of these
Hilbert spaces is superselected. Hence, we can focus on one at a
time and find all eigenvalues and eigenstates of the Hamiltonian
in it.

\subsubsection{The case $x_o =0$}
\label{s4.2.1}

Let us consider the $x_o =0$ case first. If $E$ is to be an
eigenvalue of $\hat{H}$, the coefficients $\Psi_0^{(m)}$ must fall
off sufficiently fast for $\ket{\Psi_{0}}$ to be normalizable. It
turns out that the simplest way to get a control on this fall-off
is to make a `Fourier transform' and go to the momentum
representation. Recall from section \ref{s3.2} that for each real
number $k$, there is an element $(k|$ of $\cyl^\star$ defined by:
$(k\ket{x} = e^{-ikx}$. Given any energy eigenstate $\ket{\Psi_0}
\in \HP^0$, we can evaluate its `Fourier transform'
\be \label{ft1} \psi(k) := (k\ket{\Psi_0} = \sum_{m=
-\infty}^{\infty} \Psi_0^{(m)} \, e^{-ikm\mu_o}\,  \ee
where $k \in (-\frac{\pi}{\mu_o}, \frac{\pi}{\mu_o})$; by
construction $\psi(k)$ is periodic. The eigenvalue equation
(\ref{ev2}) now becomes
\begin{equation}\label{ev3}
\frac{d^2 \psi_{0}(k)}{dk^2} + 2d^2 \left(\frac{E}{\hbar\omega} +
\frac{d^2}{\mu_o^2}\biggl[\cos(k\mu_o) -1 \biggr] \right)
\psi_{n}(k) = 0 \, .
\end{equation}
Thus, the difference equation (\ref{ev1}) in the position space
becomes a \emph{differential equation} in the momentum space.
Setting
\begin{equation} \label{eq:hbdef}
   \phi:= \frac{k\mu_o +{\pi}}{2}, \quad
 h := \frac{4d^2}{\mu_o^2}, \quad \mbox{and} \quad b:= h \cdot
\frac{2E}{\hbar \omega},
\end{equation}
the equation simplifies to:
\begin{equation}\label{mathieu}
\frac{d^2 \psi_{0}(\phi)}{d\phi^2} + \left(b - h^2 \cos^2(\phi)
\right) \psi_0(\phi) = 0 \, .
\end{equation}
This is precisely the well-studied Mathieu's equation. {}From
basic theory of differential equations we conclude
that~(\ref{mathieu}) does admit solutions. However, since the
Fourier transforms~(\ref{ft1}) of states in the position
representation are necessarily periodic, the question is whether
the solutions $\psi_0(\phi)$ are \emph{periodic} (with period
$\pi$). If they are, we may take the inverse Fourier transform and
recover a state $\ket{\Psi_0}\in \HP^0$; by Parseval's theorem
this state must be normalizable. Thus, the question of whether
$\psi_0^{(m)}$ have appropriate fall-off reduces to whether
solutions $\psi_0(\phi)$ to Mathieu's equation are periodic.

We can now appeal to the general theory of ordinary differential
equations with periodic coefficients---specifically, Floquet's
theorem---to conclude that: i) there is a discrete infinity of
periodic solutions with the required period $\pi$; and ii) each of
the corresponding energy eigenvalues is non-degenerate. (See
\cite{ince} for the general theory; \cite{as} for application to
Mathieu's equation.) Let us denote the allowed eigenvalues by
$E_{0,n}$ and the corresponding eigenstates in $\HP^0$ by
$\ket{\Psi_{0,n}} = \sum \Psi_{0,n}^{(m)} \ket{m\mu_o}$. The
question now is how these eigenvalues and eigenstates are related
to those of the Schr\"odinger theory.

By examining our definition of parameters in~(\ref{eq:hbdef}), we
see that the ratio of $\mu_o/d$ in which we are interested
corresponds to very large values of $h$. We can then employ an
asymptotic formula \cite{as} for the $b$ coefficients:
\begin{equation}
  \label{eq:bnapprox}
  b \sim (2n+1)h -\frac{2n^{2}+2n+1}{4} + \mathcal{O}(h^{-1}).
\end{equation}
By substituting this expansion back into our definition
(\ref{eq:hbdef}) of the $b$ coefficients we obtain the following
expansion for the energy eigenvalues $E_{0,n}$:
\begin{equation}
  \label{eq:maten}
  E_{n} \sim (2n+1)\frac{\hbar\omega}{2}
  -\frac{2n^{2}+2n+1}{16}\left(\frac{\mu_o}{d}\right)^{2}
  \frac{\hbar\omega}{2}
  + \mathcal{O}\!\left(\frac{\mu_o^{4}}{d^{4}}\right).
\end{equation}
Thus, in the limit $\mu_o/d \rightarrow 0$, the $E_n$ reduce to
the Schr\"odinger eigenvalues, but for the `physical' nonzero
value of $\mu_o/d$, there is a correction introduced by the
`fundamental' discreteness. We see from this equation that the
first correction to the eigenvalue is negative and of order
$\mu_o^{2}/d^{2}$. Using the very conservative value $10^{-19}$m
of $\mu_o$, for a carbon monoxide molecule we conclude that these
corrections are significant only when $n \approx 10^{7}$, i.e.,
when the vibrational energy of the oscillator is $\approx
10\,$MeV, or, in classical terms, the average vibrational velocity
is $10^{14}\,\mathrm{m s}^{-1}$. Thus, while the corrections are
conceptually important, in the domain of validity of
non-relativistic quantum mechanics they are too small to have been
observed.

Next, let us compare the eigenstate $\ket{\Psi_{0,n}}$ with the
shadow $\ket{\Psi_{0,n}^{\mathrm{shad}}}$ of the $n$th
Schr\"odinger eigenstate on the graph $\alpha^{0}$. Unfortunately,
we cannot carry out this task analytically because closed form
expressions for the Mathieu functions are not available.
Therefore, let us calculate the norm of
\begin{equation}
  \label{eq:bigddef}
  \ket{\Delta\Psi_{0,n}} := \ket{\Psi_{0,n}^{\mathrm{shad}}} -
  \ket{\Psi_{0,n}}
\end{equation}
numerically and study its behavior as a function of $n$ and
$\mu_o/d$. It turns out \cite{jw} that the log-log plot of the
norm of $\ket{\Delta\Psi_{0,n}}$ against $\mu_o$ is linear for
$10^{-6} < (\ell/d) < 1$ and $n \leq 10$, a range of parameters
for which the computation can be readily performed. By a least
squares analysis we can then verify that:
\begin{equation}
  \label{eq:normfit}
  \ip{\Delta\Psi_{0,n}}{\Delta\Psi_{0,n}}^{\frac{1}{2}} \sim
  (n+1)^{1.35}\,\left(\frac{\mu_o}{d}\right)^{1.10},
\end{equation}
These numerical results together with the analytic knowledge that
the difference equation~(\ref{ev1}) with which we began is itself
a standard discretization of Hermite's equation strongly suggest
that for $n \ll 10^{7}$, the exact eigenstates are experimentally
indistinguishable from the shadows of the Schr\"odinger
eigenstates on the graph $\alpha^0$. However, since this evidence
is not as mathematically clear-cut as other results of this paper,
we will examine this issue from a different angle in sub-section
\ref{s5.3}.

\subsubsection{General $x_o$}

Let us now consider the energy eigenstates in the Hilbert space
$\H^{x_o}$ for a general value of $x_o \in (0, \mu_0)$. Now, the
momentum wave function is given by
\be \psi_{x_o}(k) :=  (k\ket{\Psi_{x_o}}\sum_{m=
-\infty}^{\infty}\, \Psi^{(m)}_{x_o}\, e^{-ikx_o}\,
e^{ikm\mu_o}\ee
Thus the momentum space wave function $\psi_{x_o}(k)$ is no longer
periodic in $k$ but satisfies instead:
\be\label{bc} \psi_{x_o}\left(\frac{\pi}{\mu_o}\right) = e^{2\pi i x_o}
\psi_{x_o}\left(-\frac{\pi}{\mu_o}\right)\, .\ee
The differential equation that an energy eigenstate must satisfy
continues to be (\ref{ev3}) and by simple redefinitions of
parameters we are again led to the standard Mathieu equation.
Thus, the only difference between the $x_o=0$ and $x_o \not= 0$
cases lies in the boundary conditions that the solutions are to
satisfy. Again, thanks to the very exhaustive literature available
on Mathieu's equation \cite{as,ince}%
\footnote{See also \cite{mcl}, especially chapter IV and the graphs and
  accompanying discussions on pages 40 and 97--98.}
, we conclude that: i) there
is a discrete infinity of solutions satisfying (\ref{bc});
ii) each of the corresponding energy eigenvalues is
non-degenerate; and iii) the eigenvalues are very close to those
in the Schr\"odinger theory with corrections which become $O(1)$
only when energy levels corresponding to $n \approx 10^7$.

To summarize, the full polymer Hilbert space $\HP$ can be
decomposed in to orthogonal, separable subspaces $\HP^{x_o}$, each
left invariant by the algebra of observables. The energy
eigenvalue equation can therefore be solved on these subspaces
independently. In all cases, there is a discrete infinity of
eigenvalues; they are very close to the eigenvalues of the
Schr\"odinger theory in its domain of validity; and each
eigenvalue is non-degenerate. There is numerical evidence that the
eigenstates in $\HP^{x_o}$ are very close to the shadows of the
Schr\"odinger eigenstates (which naturally belong to $\cyl^\star$)
on graphs $\alpha^{x_o}$.

\emph{Remarks}: 1. Recall that for the kinetic energy term
$\hat{H}_{\mathrm{kin}}$ in the Hamiltonian, we used the operator
$\Khl$ of (\ref{h}) rather than the square $\hat{K}_{\mu_o}^2$ of
the operator $\hat{K}_{\mu_o}$ of (\ref{K}). Both alternatives
appear to be viable from the classical standpoint. However, had we
chosen $\hat{K}_{\mu_o}^2$ in place of $\Khl$, we would have found
a 2-fold degeneracy in the eigenvectors irrespective of how small
$n$ is because, in effect, the coefficients $\psi_{x_o}^{(m)}$ for
even and odd $m$ would have decoupled in the eigenvalue equation.
Hence, from a quantum mechanical perspective, only the choice
$\Khl$ is experimentally viable. This situation is familiar from
lattice gauge theories but brings out the fact that the polymer
framework has to be set up rather delicately; small $\mu_o/d$ does
\textit{not} automatically ensure that the polymer
results would be close to the continuum ones.\\
\indent 2. While all `low lying' eigenvalues are very close to
$\hbar\omega (n + \frac{1}{2})$, eigenvalues in different sectors
$\HP^{x_o}$ differ from one another slightly. Suppose for a moment
that the only limitation of Schr\"odinger quantum mechanics comes
from the fact that it ignores the inherent discreteness implied by
quantum geometry. Since the polymer particle model accounts for
this discreteness, it would then be the `fundamental' theory
underlying Schr\"odinger quantum mechanics. Then, we would be led
to conclude that the detailed energy levels of physical harmonic
oscillators would be sensitive to the physical, underlying quantum
geometry; i.e. depend on the graph which best captures the
fundamental discreteness along the line of motion of the oscillator.\\
\indent 3. Our construction was motivated by the way the
Hamiltonian constraint is treated in full general relativity. Note
however, a qualitative difference. Solutions to the Hamiltonian
constraints fail to belong to the polymer Hilbert space of quantum
geometry because zero fails to be a discrete eigenvalue of those
operators.  In the case of a harmonic oscillator, by contrast, the
full spectrum is discrete. Therefore, now the eigenvectors belong
to the polymer Hilbert space and $\cyl^\star$ is relevant only in
making contact with the Schr\"odinger quantum mechanics. Had we
considered a free particle instead, as in the Schr\"odinger
theory, the spectrum of the polymer Hamiltonian operator
$$\hat{H}_{\rm Free} = (\hbar^2/2m)\, \widehat{K_{\mu_o}^2}$$
would have been continuous. The eigenvectors would no longer be
normalizable in $\HP$ but belong to $\cyl^\star$. For energies $E
\ll \hbar^2/(2m\mu_o^2)$, they are practically indistinguishable
from plane waves in the sense that their shadows on sufficiently
refined regular graphs (with separation $\sim \mu_o$) are very
close to those of plane waves. However, as one would expect, for
higher energies, the `fundamental description' introduces major
corrections.

\subsection{Shadows of Schr\"odinger eigenstates}
\label{s5.3}

In this sub-section we will further explore the relation between
the polymer and Schr\"odiger energy eigenstates. For definiteness
let us restrict ourselves to the $x_o =0$ case, i.e., to the
Hilbert space $\HP^{0}$ associated with the graph $\alpha^{0}$.

The shadows on $\alpha^{0}$ of the Schr\"odinger eigenstates
$\brar{\Psi_{n}}$ are given by
\be  \label{eq:hshonshad}
  \ket{\Psi^{\mathrm{shad}}_{0,n}} =
  c_{n}\,\sum_{m}\,H_{n}\!\left(\frac{m{\mu_o}}{d}\right)\,
  e^{-\textstyle{\frac{m^{2}{\mu_0}^{2}}{2d^{2}}}} \,\ket{m{\mu_o}}.
\ee
where $c_{n}$ is the normalization constant. The main result of
this sub-section is that these shadows satisfy the eigenvalue
equation of the polymer Hamiltonian (\ref{h}) to an excellent
degree of approximation if $n \ll 10^7$.

The action of the Hamiltonian (\ref{h}) on an arbitrary state
$\ket{\Psi} = \sum_m \psi(m) \ket{m{\mu_o}}$ can be easily
calculated.  The result is:
\begin{equation}
   \hat{H} \, \ket{\Psi} = \frac{\hbar \omega d^2}{2{\mu_o}^2}
   \sum_m \left[ \left(2 + \frac{m^2{\mu_o}^4}{d^4}\right) \psi(m) -
   \biggl(\psi(m+1) + \psi(m-1) \biggr) \right]\,\ket{m{\mu_o}}\,.
\end{equation}
Let us begin with the shadow ground state
$\ket{\Psi^{\mathrm{shad}}_{0,0}}$. We have:
\begin{equation}
\label{hon0shad}
   \hat{H}\,\ket{\Psi_{0,0}^{\mathrm{shad}}} = \frac{\hbar \omega
     d^2}{2{\mu_o}^2}\,c_{0}\,
        \sum_m e^{- \textstyle{\frac{m^2 {\mu_o}^2}{2 d^2}}}
     \left[\left(2 + \frac{m^2{\mu_o}^4}{d^4}\right) -
        e^{-\textstyle{\frac{{\mu_o}^2}{2d^2}}}
        \biggl(e^{-\textstyle{\frac{m{\mu_o}^2}{d^2}}}
        + e^{\textstyle{\frac{m{\mu_o}^2}{d^2}}}
        \biggr) \right]\, \ket{m{\mu_o}}\, .
\end{equation}
To make the structure of the right side more transparent, let us
expand the last three exponentials and keep the lowest few terms:
\begin{equation}
\label{psi0harm}
\begin{split}
   \hat{H}\,\ket{\Psi_{0,0}^{\mathrm{shad}}} &= \frac{\hbar \omega
     d^2}{2{\mu_o}^2}\,c_{0}\,
        \sum_m e^{-\textstyle{\frac{m^2 {\mu_o}^2}{2 d^2}}} \,\times \\
   & \qquad \left[\left(2 + \frac{m^2{\mu_o}^4}{d^4}\right) -
        \left(2 - \frac{{\mu_o}^2}{d^2} + \frac{{\mu_o}^4}{4d^4} +
        \frac{m^2{\mu_o}^4}{d^4} - \frac{{\mu_o}^6}{24d^6} -
        \frac{m^2{\mu_o}^6}{2d^6} + \cdots \right) \right]  \,
        \ket{m{\mu_o}}  \\
   &= \frac{\hbar\omega}{2}\,\ket{\Psi_{0,{\mu_o}}^{\mathrm{shad}}} +
         \frac{\hbar\omega}{2}\,\frac{c_{0}}{4}\,
         \sum_m e^{-\textstyle{\frac{m^2 {\mu_o}^2}{2
         d^2}}}\,\left[ - \frac{{\mu_o}^2}{d^2} +
           \frac{{\mu_o}^4}{6d^4} + \frac{2m^{2}{\mu_o}^{4}}{d^{4}} -
           \cdots\right]\,\ket{m{\mu_o}} \, .
\end{split}
\end{equation}
Thus, we have
\be \label{eq:hshoonshad}
\hat{H}\,\ket{\Psi_{0,0}^{\mathrm{shad}}} = \frac{\hbar\omega}{2}
\,\left[\ket{\Psi_{0,0}^{\mathrm{shad}}} +
\ket{\delta\Psi_{0,0}}\right] \ee
where $\ket{\delta\Psi_{0}}$ is $c_{0}/4$ times the last sum
in~(\ref{psi0harm}). Since this `remainder' proportional to
${\mu_o}^{2}/d^{2}$, in the series in square brackets only terms
with large $m$ make significant contributions and these terms are
severely damped by the exponential multiplicative factor. Hence it
is plausible that $\ip{\delta\Psi_{0}}{\delta\Psi_{0}} \ll 1$,
i.e., that the shadow state is very nearly an eigenstate of
$\hat{H}$. We will first establish that the situation is similar
for all excited states and then show that the expectation on
smallness of the remainder term is correct \textit{for all}
eigenstates.

Let us then act on the shadow (\ref{eq:hshonshad}) of the $n$-th
excited state with the Hamiltonian. We obtain:
\begin{multline}
  \label{eq:honnshad}
  \hat{H}\,\ket{\Psi^{\mathrm{shad}}_{0,n}} = \frac{\hbar\omega
    d^{2}}{2{\mu_o}^{2}}c_{n}\,\sum_{m}\,
  e^{-\frac{m^{2}{\mu_o}^{2}}{2d^{2}}}\,\left\{ \left(2 +
        \frac{4m^{2}{\mu_o}^{4}}{d^{4}}\right)\,H_{n}
        \!\left(\frac{m{\mu_o}}{d}\right)
        -e^{-\frac{{\mu_o}^{2}}{2d^{2}}}\times\right. \\
        \left.\left[e^{-\frac{m{\mu_o}^{2}}{d^{2}}}\,H_{n}
        \!\left(\frac{m{\mu_o}}{d}+
        \frac{{\mu_o}}{d}\right) + e^{\frac{m{\mu_o}^{2}}{d^{2}}}\,H_{n}\!
        \left(\frac{m{\mu_o}}{d}-\frac{{\mu_o}}{d}\right)\right]\right\}
\end{multline}

This expression can be simplified using the basic recurrences
satisfied by the Hermite polynomials and by expanding the
exponentials using Taylor's theorem. As with the
ground state, we can then conclude
\begin{equation}
  \hat{H}\,\ket{\Psi_{0,n}^{\mathrm{shad}}} =
  \frac{2n+1}{2}\,\hbar\omega\,\ket{\Psi_{0,n}^{\mathrm{shad}}} +
  \frac{\hbar\omega}{2}\,\ket{\delta\Psi_{n}}
\end{equation}
where $\ket{\delta\Psi_{n}}$ can be evaluated explicitly. To bound
its norm we use the fact that the sums appearing in the norm are
Riemann sums for integrals that may be evaluated analytically. In
this way it is possible to prove that:
\begin{equation}
  \label{eq:normbound}
\ip{\delta\Psi_{n,{\mu_o}}}{\delta\Psi_{n,{\mu_o}}}^{\frac{1}{2}}
  \,=\,
  \frac{\sqrt{35}}{48}\left(2n^{4}+4n^{3}+10n^{2}+8n+3\right)^{\frac{1}{2}}
   \left(\frac{{\mu_o}}{d}\right)^{2}
   +\mathcal{O}\!\left(n^{3}\left(\frac{{\mu_o}}{d}\right)^{4}\right).
\end{equation}

We see immediately  that
$\ip{\delta\Psi_{n}}{\delta\Psi_{n}}^{1/2}$ approaches zero if we
let ${\mu_o}/d$ approach zero. For finite ${\mu_o}/d$, its value
depends on $n$ and, as one would expect on physical grounds, is of
order unity when $n \sim d/{\mu_o}$. In the case of the molecular
vibrations of carbon monoxide considered above, this corresponds
to $n\sim 10^{7}$.  It is obvious that (among other things) the
approximation that $V(x)$ can be described by the simple harmonic
oscillator potential will break down long before this energy level
$n$ is reached.

For the bound~(\ref{eq:normbound}) to be useful, we must know when
the $\mathcal{O}(n^{3}\,({\mu_o}/d)^{4})$ term is negligible. This
can easily be investigated numerically, and it is found that the
asymptotic behavior~(\ref{eq:normbound}) is attained almost as
soon as ${\mu_o}/d < 1$.  To give some examples, one finds that
for the ground state, even when ${\mu_o}/d$ is as large as 0.1,
equation~(\ref{eq:normbound}) is accurate to less than a percent
and the magnitude of the norm of $\ket{\delta\Psi_{n}}$ is about
$2.2 \times 10^{-3}$. For $n=9$ and ${\mu_o}/d = 10^{-3}$,
equation~(\ref{eq:normbound}) is accurate to one part in $3\times
10^{-5}$ when ${\mu_o}/d = 10^{-3}$, and the magnitude of the norm
of $\ket{\delta\Psi_{n}}$ is $2.13\times10^{-13}$. Thus, not only
does the norm of $\ket{\delta\Psi_{n}}$ approach zero as
${\mu_o}/d$ approaches zero, but it also quickly approaches the
asymptotic behavior of equation~(\ref{eq:normbound}).

To summarize, we have shown that the shadows of the Schr\"odinger
energy eigenstates on the graph $\alpha^0$ are eigenstates of the
polymer Hamiltonian $\hat{H}$ to a high degree of approximation at
`low' energies. Quantum geometry effects manifest themselves only
at energy levels as high as $n \sim 10^7$, i.e., long beyond the
validity of non-relativistic approximation. This result
complements our findings in Section \ref{s5.2} where we compared
the exact eigenstates of the polymer Hamiltonian with the shadows
of the Schr\"odinger Hamiltonian.

\section{Discussion}
\label{s7}

We began, in section \ref{s1}, by raising three conceptual issues
of a rather general nature that arise in relating background
independent approaches to quantum gravity with low energy physics:
i) What is the precise sense in which semi-classical states arise
in the full theory? ii) Is the fundamental Planck scale theory,
with an in-built fundamental discreteness, capable of describing
also the low energy physics rooted in the continuum, or, does it
only describe an entirely distinct phase?  iii) Can one hope to
probe semi-classical issues without a canonical inner product on
the space of physical states $\cylstar$? To probe these issues in
a technically simpler context, we introduced the `polymer
framework' in a toy model---a non-relativistic particle---where
the same questions arise naturally. In the context of the model,
we found encouraging answers to all three questions: although at
first the polymer description seems far removed from the standard
Schr\"odinger one, the second can be recovered from the former in
detail.

Specifically, we could: a) give a criterion to select the coherent
states entirely within the polymer framework and, using their
shadows, demonstrate in detail that they are sharply peaked about
the corresponding classical states; and, b) introduce the
Hamiltonian operators in the polymer framework and show that their
eigenvalues and eigenfunctions are indistinguishable from those of
the continuum, Schr\"odinger theory within its domain of validity.
Logically, one can forego the continuum theory, work entirely with
the polymer description, and compare the theoretical predictions
with experimental results. However, since we already know that
Schr\"odinger quantum mechanics reproduces the experimental
results within its domain of validity, it is simpler in practice
to verify agreement with the Schr\"odinger results. As one might
physically expect, since the polymer framework `knows' about the
underlying discreteness, it predicts corrections to the
Schr\"odinger framework which become significant when the energies
involved are sufficiently high to probe that discreteness. Thus,
we have a concrete mathematical model, inspired by loop quantum
gravity, which realizes the idea that a fundamental theory can be
radically different from the continuum theory \emph{both
conceptually and technically} and yet reproduce the familiar low
energy results.

The broad strategy we followed, including the use of shadow
states, was already outlined in the general program \cite{al7}.
Notions needed in this analysis are all available in field
theories as well as full quantum gravity. The details of the
polymer particle toy model provide concrete hints for these more
complicated theories. For example, it may seem `obvious' in that
calculations on sufficiently refined graphs should reproduce the
continuum answers. Our analysis showed that this is not
necessarily the case. Naively, one would have used the operator
$(\hbar^2/2m)\hat{K}^2$ as the kinetic part of the quantum
Hamiltonian. However, this choice would have given a two-fold
degeneracy for all eigenstates of the polymer Hamiltonian
including the `low energy' ones, while in the Schr\"odinger
theory, all eigenstates are non-degenerate. This is a concrete
illustration of how the requirement that the theory should
reproduce predictions of well-established theories in the
\emph{low energy regime} can be used to discriminate between
choices available in the construction of the `fundamental'
framework. A second example arises from a cursory examination of
the form of the polymer particle Hamiltonian. While the potential
continues to be unbounded as in the Schr\"odinger theory, the
kinetic part of the Hamiltonian is now bounded. Therefore, at
first, it seems that the kinetic term will not be able to `catch
up' for large $x$ to produce normalizable solutions to the
eigenvalue equation $((\hbar^2/2m) \Khl + x^2 )\ket{\Psi} = E
\ket{\Psi}$. Furthermore, this expectation can be `confirmed' by
numerical solutions to the difference eigenvalue equation
(\ref{ev2}). However the careful examination of section
\ref{s5.2}, involving the Mathieu equation in the momentum
representation, showed that the expectation is incorrect and the
divergence of $\ket{\Psi}$ one encounters in computer calculations
is just a numerical artifact. Finally, even at the kinematical
level, there is a subtlety: a priori, it is not at all obvious
that any calculation to select semi-classical states $\brar{\Psi}$
in $\cylstart$, carried out entirely within the polymer framework,
will reproduce the \emph{standard} coherent states. One could
indeed be working in an inequivalent `phase' of the theory and
thus find that there are no semi-classical states at all or
discover states which are semi-classical in a certain well-defined
sense but distinct from the standard coherent states (as in, e.g.,
\cite{toh}). The fact that the Schr\"odinger semi-classical states
can be recovered in the polymer framework is thus non-trivial and
suggests how \emph{standard} low energy physics could emerge from
the polymer framework. Thus, our analysis provides useful
guidelines for more realistic theories, pointing out potential
pitfalls where care is needed and suggesting technical strategies.

Finally, there are also some conceptual lessons. First, we saw
concretely that recovery of semi-classical physics entails two
things: isolation of suitable states \emph{and} a suitable coarse
graining. In the toy model, the coarse graining scale was set by
our tolerance $d$ and continuum physics emerges only when we
coarse grain on this scale. A second lesson is that the
availability of a scalar product on the space of physical states
is not essential at least for semi-classical considerations: The
framework of shadow states---with its Wilsonian
overtones---provides an effective strategy to recover low energy
physics. In the next paper in this series we will see that these
ideas can be naturally elevated to the quantum Maxwell theory.

\begin{acknowledgments}
We would like to thank Amit Ghosh and participants of an Erwin
Schr\"odinger Institute workshop for stimulating discussions. We
are especially grateful to Detlev Buchholz, Klaus Fredenhagen,
Rodolfo Gambini, Chris Fewster, Stefan Hollands, Jerzy
Lewandowski, Hanno Sahlmann, Thomas Thiemann, Bill Unruh, Robert
Wald and Jacob Yngvasson for raising issues which significantly
clarified the final presentation. This work was supported in part
by the NSF grant PHY-0090091 and the Eberly research funds of Penn
State. In addition, SF was supported by the Killam Trust at the
University of Alberta, and JLW, through the Duncan Fellowship of
Penn State.
\end{acknowledgments}
\bigskip

\appendix

\section{The displacement operator $\hat{V}(\mu)$ and holonomies}
\label{a1}

Recall that the displacement operators $\hat{V}(\mu)$ are the
analogs of holonomy operators in Maxwell theory and quantum
geometry. In this Appendix, we collect some properties of
displacement operators which will be useful in the discussion of
holonomies in the subsequent papers.

We begin by recalling the commutator between $\hat{x}$ and
$\hat{V}(\mu)$:
\begin{equation}\label{xvcommapp}
   [\hat{x}, \hat{V}(\mu)] = - \mu \hat{V}(\mu) \, .
\end{equation}
This equation gives rise to interesting uncertainty relations,
even though $\hat{V}(\mu)$ are unitary rather than self-adjoint
\cite{ll}. To obtain these, let us decompose $\hat{V}$ into the
sum of two Hermitian operators,
\begin{equation}\label{defcs}
   \hat{V}(\mu) = \hat{C}(\mu) + i \hat{S}(\mu) \, .
\end{equation}
It is straightforward to obtain the commutation relations between
$\hat{C}$, $\hat{S}$ and $\hat{x}$ from (\ref{xvcommapp}) as
\[ [\hat{x} , \hat{C}(\mu)] = - i\mu \hat{S}(\mu) \qquad \mbox{and}
   \qquad [\hat{x} , \hat{S}(\mu)] = i\mu \hat{C}(\mu) \, .
\]
Therefore, we can obtain uncertainty relations between $\hat{C}$,
$\hat{S}$ and $\hat{x}$:
\begin{equation}\label{uncertcs}
   (\Delta x)^2 (\Delta C(\mu))^2 \ge \frac{\mu^2}{4}
   \langle\hat{S}(\mu)\rangle^2
   \qquad \mbox{and} \qquad  (\Delta x)^2 (\Delta S(\mu))^2 \ge
   \frac{\mu^2}{4} \langle\hat{C}(\mu)\rangle^2
\end{equation}
Now, it is natural to define the uncertainty in $\hat{V}$ as
\begin{equation}\label{deltav}
   (\Delta V)^2 := \langle V^{\dagger} V\rangle - |\langle V \rangle|^2
   = 1 - (\langle C \rangle^2 + \langle S\rangle^2) \, ,
\end{equation}
where the second expression follows from the unitarity of
$\hat{V}$ and the definitions of $\hat{C}$ and $\hat{S}$
(\ref{defcs}). Finally, combining (\ref{deltav}) and
(\ref{uncertcs}) we obtain the desired uncertainty relation
\begin{equation}\label{uncert}
   (\Delta x)^2 \frac{(\Delta V(\mu))^2}{1-(\Delta V(\mu))^2}  \ge
   \frac{\mu^2}{4} \, .
\end{equation}

It is natural to ask how close the semi-classical states of
section \ref{s4.1} come to saturating this bound. Let us begin by
considering the state $\brar{\Psi_{o}}$ peaked at ($x$=0, $k$=0).
The `expectation value' of $\hat{V}(\mu)$ in $\brar{\Psi_{o}}$ and
its shadow $\ket{\Psi_{o,\ell}^{\rm shad}}$ on a regular lattice
with spacing $\ell$ is given in (\ref{vmu}) as
\[
    \langle \hat{V}(\mu) \rangle  \approx
    e^{-\textstyle{\frac{\mu^2}{4d^2}}} \left( 1 +
    e^{-\textstyle{\frac{\pi^{2}d^2}{\ell^{2}}}}
    \left[2\cos\left(\frac{\pi\mu}{\ell}\right) - 2
    \right] \right) \, .
\]
Then, it is straightforward to evaluate the fluctuations of
$\hat{V}(\mu)$ as
\begin{equation} \label{deltav2}
   (\Delta \hat{V}(\mu))^2 := 1 - |\evalue{\hat{V}(\mu)}|^2 \approx 1 -
    e^{-\textstyle\frac{\mu^2}{2d^2}} \, ,
\end{equation}
where we have neglected corrections of order $\exp
(-\textstyle{\frac{\pi^2 d^2}{l^2}})$. Combining (\ref{deltav2})
with the fluctuations in $x$ (\ref{deltax}), we obtain:
\begin{equation}
   (\Delta x)^2 \cdot \frac{(\Delta V(\mu))^2}{1-(\Delta V(\mu))^2}
   \approx \left(\frac{d^2}{2} \right) \cdot
   \left(\frac{1 - e^{-\textstyle{\frac{\mu^2}{2d^2}}}}
   {e^{-\textstyle{\frac{\mu^2}{2d^2}}}}\right)\, .
\end{equation}
Thus, for a general $\mu$, we are not close to saturation.
However, if $\mu \ll d$, we can expand in powers of $\mu /
d$ to obtain:
\begin{equation}\label{uncertxv}
   (\Delta x)^2 \frac{(\Delta V(\mu))^2}{1-(\Delta V(\mu))^2} =
   \left(\frac{\mu^2}{4}\right)\left(1 +
   \mathcal{O}\!\left(\frac{\mu^2}{d^2}\right) \right) \, .
\end{equation}
Thus, in this case, the uncertainty relation (\ref{uncert}) is
indeed saturated, modulo terms of the order $(\mu/d)^2$. If $\mu
\sim \ell$, a similar result can be obtained for general coherent
states peaked at any value of momentum $k$, \emph{even when $k$
approaches $\pi / \ell$}.  Note that this in marked contrast to
the uncertainty relation between $\hat{x}$ and $\hat{K}_{\mu_{o}}$
which is similarly saturated only if $k \mu_{o} \ll 1$.

Finally, a natural question is whether the `expectation value' of
$\hat{V}(\mu)$ can be used to determine the momentum of the system
when it is in a semi-classical state. In a semi-classical state
labelled by $\zeta= \textstyle{\frac{1}{\sqrt{2} d}}(x+ id^2k)$, the
`expectation value' of $\hat{V}(\mu)$ is given by
\begin{equation}
   \evalue{\hat{V}(\mu)} = e^{-\textstyle\frac{\mu^2}{4d^2}}\,
   e^{-ik\mu}\,\left[ 1 +
     \mathcal{O}\!\left(e^{-\textstyle{\frac{\pi^2
             d^2}{\ell^2}}}\right)\right] \, .
\end{equation}
An obvious strategy is to just define the `expected momentum'
$\tilde{k}$ in the quantum state $\brar{\Psi_\zeta}$ to be:
\begin{equation}\label{p}
  \evalue{\hat{V}(\mu)} = |\evalue{\hat{V}(\mu)}|\,\, e^{-i\mu \,
  \tilde{k}}
  \, ,
\end{equation}
i.e., to associate the momentum $\tilde{k}$ with the phase of the
$\hat{V}$ operator. Clearly, modulo corrections $\mathcal{O}
(\exp\, -{\pi^2 d^2}/{\ell^2})$, $\tilde{k}$ equals $k$. Moreover,
this result holds even if $k \sim \pi / \ell$. The
$|\evalue{\hat{V}(\mu)}|$ factor in our expression (\ref{p}) may
seem surprising. However, it does \textit{not} arise because of
the polymer nature of the Hilbert space we are considering; it is
necessary also in the Schr\"odinger representation. Note also that
our expression (\ref{deltav2}) for the variation of $V$, implies
that $|\evalue{\hat{V}(\mu)}|$ must be less than one if
$\mu\not=0$. Otherwise the fluctuation in $V$ will vanish, i.e.,
we will have a state of definite momentum and the uncertainty
relation (\ref{uncert}) would imply that the state must have
infinite spread in $x$.

Techniques introduced in this appendix will be useful when it
comes to examining expectation values and fluctuations of
holonomies in Maxwell and gravitational semi-classical states.

\section{Approximate consistency}
\label{a2}

In the main body of the paper we introduced `fundamental
operators' such as $\hat{K}_{\mu_o}$ and $\hat{H}$ on the entire
polymer Hilbert space $\HP$ and analyzed their properties. In
field theories, by contrast, one often ties operators to the
energy scale under consideration and constructs from them `an
effective field theory' a la Wilson. Such constructions are likely
to play an important role in relating quantum field theories on
quantum geometries with low energy physics. Therefore, in this
appendix, will extend some of the considerations of sections
\ref{s4} and \ref{s5} by allowing operators which are tied to the
lattice spacing under consideration. For example, by setting
$\hat{K}_{\ell} = (i/2\ell) (\hat{V}(\ell) - \hat{V}(-\ell))$, we
obtain a family of `momentum' operators $\hat{K}_{\ell}$, one for
each regular lattice. The dependence of such operators on $\ell$
in the limit $\ell \rightarrow 0$ will enable us to relate our
constructions to the Wilsonian renormalization group flow. We will
now examine properties of such families of operators and introduce
the notion of `approximate consistency in the low energy regime',
which will be useful in the analysis of field theories in later
papers.

As mentioned in section \ref{s2}, operators on the full Hilbert
space $\HP$ in quantum geometry often arise from \emph{consistent
families} of operators on the Hilbert spaces $\{\Hilb_{\gamma}\}$
associated to graphs $\gamma$ \cite{al3}.  However, since we will
be interested in `low energy' states that lie in $\cylstar$ but
not in $\HP$, we will use as our starting point the consistency of
families of operators on $\cylstar$.%
\footnote{We are grateful to Jurek Lewandowski for pointing out
the utility of this definition.}
This concept is defined naturally using the duality between $\cyl$
and $\cylstar$.  Specifically, if we are given a family of
operators $\{\hat{O}_{\gamma}\}$ defined on each Hilbert space
$\Hilb_{\gamma}$, then this family is said to be
\textit{consistent on $\cylstar$} if, given any state
$\brar{\Psi}\,\in\,\cylstar$, any two graphs $\gamma$ and
$\gamma'$ such that $\gamma\,\subseteq\,\gamma'$, and any state
$\ket{\phi_{\gamma}}\,\in\,\cyl_{\gamma}$, the following holds:
\begin{equation}
  \label{eq:15}
  \rexpect{\Psi}{\hat{O}_{\gamma}}{\phi_{\gamma}} =
  \rexpect{\Psi}{\hat{O}_{\gamma'}\Pi^{*}_{\gamma\gamma'}}
  {\phi_{\gamma}}\, . \end{equation}
Here $\Pi^{*}_{\gamma\gamma'}$ denotes the pull-back from
$\cyl_{\gamma}$ to the larger Hilbert space $\cyl_{\gamma'}$. This
condition serves to ensure that the matrix elements of the
operator are independent of the graph $\gamma$ used to calculate
them, i.e., that there is a single operator $\hat{O}$ on
$\cylstart$ such that $\rexpect{\Psi}{\hat{O}}{\phi_{\gamma}} =
\rexpect{\Psi}{\hat{O}_{\gamma}}{\phi_{\gamma}}$ for all graphs
$\gamma$. In the polymer particle example, several important
operators are consistent on $\cylstar$, including the position
operator $\hat{x}$ and the displacement operator $\hat{V}(\mu)$.

However, the new families of operators such as $\hat{K}_\ell$,
defined above, do \textit{not} form a consistent family.  Neither
would be the family of Hamiltonian operators $\hat{H}_{\ell}$ on
$\H_{\ell}$, if their definitions were similarly tied to the
lattice spacing. (The Hamiltonians defined in lattice gauge theory
are typically of this type.) To examine such families of
operators, we must weaken our definition of consistency on
$\cylstar$.

We do so in two directions.  First, since the momentum operators
are intimately connected to differentiation, we cannot expect a
weakened form of~(\ref{eq:15}) to hold for arbitrary states in
$\cylstart$, but only for `low energy ones', i.e., states that are
elements of $\mathcal{S}$. Second, we do not require expectation
values in~(\ref{eq:15}) to be exactly equal, but instead only that
the norm of their difference should be small. Finally, as in the
main text, we will only consider regular lattices.  We then say
that a family of operators $\{\hat{O}_{\gamma}\}$ defined on
\textit{regular} lattices $\gamma$ is \textit{approximately
consistent on low energy states} if, given a constant $o_{0}$
(with same dimensions as $\hat{O}$), an $\epsilon\,>\,0$, and any
two states $\psi(x),\,\phi(x)\,\in\,\mathcal{S}$, there exists a
regular lattice $\gamma$ such that for any regular lattice
$\gamma'$ that is a refinement of $\gamma$, the following holds:
\begin{equation}
  \label{eq:16}
  \frac{|\rexpect{\Psi}{\hat{O}_{\gamma}}{\Phi^{\mathrm{shad}}_{\gamma}} -
   \rexpect{\Psi}{\hat{O}_{\gamma'}\Pi^{*}_{\gamma\gamma'}}
    {\Phi^{\mathrm{shad}}_{\gamma}}|}{\norm{\Psi^{\mathrm{shad}}_{\gamma}}\,
    \norm{\Phi^{\mathrm{shad}}_{\gamma}}}   \,\,<\,\,o_{0}\,\epsilon.
\end{equation}
Note that, in this definition, it is essential that we divide by
the appropriate norms since states in $\cylstar$ are not
normalized.

It is obvious from this definition that any consistent family of
operators is automatically approximately consistent on low energy
states. Moreover, it is not hard to show that the family of
momentum operators $\hat{K}_{\ell}$ and Hamiltonians
$\hat{H}_{\ell}$ satisfies this definition as well. The proof
follows from the fact that the two matrix elements
in~(\ref{eq:16}) (divided by the appropriate norms) are Riemann
sums for the same integrals. Hence, since they converge in the
$\ell\rightarrow 0$ limit to the same thing, they form a Cauchy
net and equation~(\ref{eq:16}) follows (see \cite{jw} for
details).

Thus, we have generalized the usual notion of consistent families
of operators to important families of operators that do
\textit{not} form a consistent family, thus allowing us to use
techniques in analyzing such operators that are similar to those
that have played such an important role in quantum geometry. This
generalization will be useful in subsequent papers on the relation
between `polymer field theories' on quantum geometry and the
familiar low energy field theories in the continuum.

\end{document}